\documentclass[fleqn,usenatbib]{mnras}

\usepackage{newtxtext,newtxmath}

\usepackage[T1]{fontenc}

\usepackage{graphicx}	
\usepackage{amsmath}	
\usepackage{comment}

\newcommand{\nc}{\newcommand}

\nc{\msun}{\ensuremath{\mathrm{M}_\odot}}
\nc{\lsun}{\ensuremath{\mathrm{L}_\odot}}
\nc{\thCO}{$^{13}$CO}
\nc{\CeiO}{C$^{18}$O}
\nc{\kms}{\mbox{km~s$^{-1}$}}
\nc{\Kkms}{\mbox{K\,km~s$^{-1}$}}

\nc{\twCO}{$^{12}$CO}
\def\O0{C$^{0}$}

\newcommand\arcdeg{\mbox{$^\circ$}}%
\nc{\cmsq}{\mbox{cm$^{-2}$}}
\nc{\cmcub}{\mbox{cm$^{-3}$}}
\nc{\tkin}{\mbox{$T_{\rm kin}$}}

\newcommand{\CII}{[C {\sc ii}]}
\newcommand{\OI}{[O {\sc i}]}
\def\ptsec{$''\mskip-7.6mu.\,$}
\def\cplus{C$^{+}$}
\nc{\thCII}{[$^{13}$C {\sc ii}]}

\title[Atomic oxygen in N2023]{Constraining the geometry of the
reflection nebula NGC\,2023 with \OI: Emission \& Absorption}

\author[B. Mookerjea et al.]{
Bhaswati Mookerjea$^{1}$\thanks{E-mail: bhaswati@tifr.res.in},
G\"oran Sandell$^{2}$,  Rolf G\"usten$^{3}$, Helmut Wiesemeyer$^{3}$, 
Yoko Okada $^{4}$, Karl Jacobs$^{4}$
\\
$^{1}$Department of Astronomy \& Astrophysics, Tata Institute of
Fundamental Research, Homi Bhabha Road, Mumbai 400005, India\\
$^{2}$ Institute for Astronomy, University of Hawai‘i at Manoa, 640 N. Aohoku Place, Hilo, HI 96720, USA\\
$^{3}$ Max Planck Institut f\"ur Radioastronomie, Auf dem H\"ugel 69, D-53121 Bonn, Germany \\
$^4$ I. Physikalisches Institut, Universit\"at zu K\"oln Z\"ulpicher Str. 77, 50937 K\"oln, Germany\\}

\date{}

\pubyear{2023}

\begin{document}
\label{firstpage}
\pagerange{\pageref{firstpage}--\pageref{lastpage}}
\maketitle

\begin{abstract}
We have mapped the  NGC\,2023 reflection nebula in the 63 and 145\,\micron\ transitions of \OI\ and the 158\,\micron\ \CII\ spectral lines using the heterodyne receiver upGREAT on SOFIA. The observations were used to identify the diffuse and dense components of the PDR traced  by the \CII\ and \OI\ emission, respectively. The velocity-resolved observations reveal the presence of a significant column of low-excitation atomic oxygen, seen in absorption in the \OI\ 63\,\micron\ spectra,  amounting to about  20--60\% of the oxygen column seen in emission in the \OI\ 145\,\micron\ spectra. Some self-absorption is also seen in \CII, but for the most part it is hardly noticeable. The \CII\  and \OI\ 63\,\micron\ spectra show  strong red- and blue-shifted wings due to photo evaporation flows especially in the southeastern and southern part of the reflection nebula, where comparison with the mid- and high-$J$ CO emission indicates that the \cplus\ region is expanding into a dense molecular cloud. Using a two-slab toy model the large-scale self-absorption seen in \OI\ 63\,\micron\ is readily explained as originating in foreground low-excitation gas associated with the source. Similar columns have also been
observed recently in other Galactic photon-dominated-regions (PDRs). These results have two implications: for the velocity-unresolved extra-galactic observations this could impact the use of  \OI\ 63\,\micron\ as a tracer of massive star formation and secondly the widespread self-absorption in \OI\ 63\,\micron\ leads to underestimate of the column density of atomic oxygen derived from this tracer and necessitates the use of alternative indirect methods.
\end{abstract}




\section{Introduction} 

The fine-structure lines of \CII\  at 158 $\mu$m and \OI\  at 63 and 145 $\mu$m
 are the  most important cooling lines in the far-infrared (FIR) and have long been used to study photon-dominated regions (PDRs).  In the rest of the paper we refer to the \OI\ 63\,\micron\ and  \OI\ 145\,\micron\  fine-structure lines simply as \OI\ 63 and \OI\ 145.  PDRs  are regions where far-ultraviolet (FUV; 6 eV $<$ h$\nu$ 13.6 eV) radiation from young massive stars dominates the physics and the chemistry of the interstellar medium and correspond to the transition from ionized to molecular gas \citep{Hollenbach1997}. PDRS were first studied extensively with the Infrared Space Observatory, though the lack of spectral and spatial resolution was a severe limitation.  This changed with the Kuiper Airborne Observatory (in limited regions) and with the Heterodyne Instrument for the Far Infrared (HIFI) on Herschel, which had both velocity resolution and sensitivity to even enable determination of the optical depth of \CII\ by also observing $^{13}$\CII\ \citep{Ossenkopf2013}. After Herschel this work was taken over by first GREAT on SOFIA and later with the array receiver upGREAT. These studies have found that the \CII\ emission often can be moderately optically thick and sometimes significantly self-absorbed \citep{Mookerjea2018, Guevara2020, Graf2012}, which may underestimate \CII\ column densities by a factor of two to three. The \OI\ 63\,\micron\ can also be optically thick and self-absorbed, sometimes even more strongly than \CII\ \citep[see e. g.,][]{Mookerjea2021}.

NGC\,2023, illuminated by the B2 star HD\,37903, is one of
the best-studied reflection nebulae. It is nearby, $\sim$400 pc, and exhibits a
strong, nearly edge-on PDR in which the cavity expands into the surrounding
dense molecular cloud. It has served as a test bed for exploring PDR models
\citep[e.g.][]{Kaufman2006}. It was mapped with GREAT on SOFIA in \CII\ and
CO(11--10) with the single pixel  L1/L2 mixer \citep{Sandell2015}, who also mapped
the nebula in $^{13}$CO(3-2), and various transitions of CO from J=3--2 up to J=7--6.

Here we have additionally mapped the NGC\,2023 PDR in the 63 and 145 $\mu$m
fine-structure transitions of \OI\ and at the same time obtained a deeper map of \CII\ using upGREAT on SOFIA. The \OI\ observations enable us for the first time to study the physical conditions in the south eastern and southern part of the nebula, where the \cplus\ region is expanding into the dense surrounding molecular cloud as well as  estimate the relative contribution of the dense and diffuse PDR gas to the \CII\ emission.

The newly obtained observations are analyzed in combination with the previously published maps of (i) \thCO(3--2), CO(6--5), CO(7--6) observed with APEX with beamsizes of 18\farcs5, 9\farcs1 and 7\farcs7 respectively and (ii) CO(11--10) observed with SOFIA/GREAT with a beamsize of 23\arcsec\ \citep{Sandell2015}.

\section{Observations}
\label{sec:observations}

The reflection nebula NGC\,2023  was observed on a 90 minute leg with the SOFIA
upGREAT\footnote{The German REceiver for
Astronomy at Terahertz frequencies (upGREAT) is a development
by the MPI f\"ur Radioastronomie and the KOSMA/Universit\"at
zu K\"oln, in cooperation with the DLR Institut f\"ur Optische
Sensorsysteme.} receiver \citep{Risacher2016} in GREAT consortium time on SOFIA flight
\#668 (Project ID: 83\_0731) out of Palmdale, CA, on March 6, 2020.  Although NGC\,2023 was observed at
low flight altitude, 11.6 km (38,000 feet), the observing conditions were very
good with a precipitable water vapor (PWV) around 5--10\,\micron. The upGREAT
was in the Low Frequency Array/High Frequency Array (LFA/HFA) configuration with
both arrays operating in parallel. The V polarization of the LFA was tuned to
\CII\ at 1.9005369 THz while the H polarization was tuned to the \OI\
145\,\micron\ line at 2.06006886\,THz. The HFA was tuned to the \OI\
63\,\micron\ at 4,744.77749\,THz. We made total power on-the fly (OTF) maps in
classical position switched mode. The reference position was at
+390\arcsec,$-$70\arcsec\ relative to HD\,37903 ( ($\alpha_{J2000}$:
$05^h~41^m~38.39^s$, $\delta_{J2000}$: $-02^\circ 15^\prime$ 32\ptsec5). The OTF
map was done in two coverages, scanning in both x and y. The map was rotated by
45\arcdeg.  The sampling was done every 3\arcsec\ with a sampling time of 0.3
second per dump. This resulted in maps for the LFA array of $\sim$ 4\farcm9
$\times$  3\farcm9, while the map size for the HFA array was 3.0\arcmin 
$\times$ 2\farcm1, which was enough to cover the SE quadrant of the reflection
nebula, where \OI\ was expected to be strong.

The observations were reduced and calibrated by the GREAT team. The GREAT team
also provided beam sizes (14\ptsec1 for \CII, 13\ptsec0 for \OI\ 145\,\micron,
and 6\ptsec3 for \OI\ 63\,\micron) and beam efficiencies derived from planet
observations. The data were corrected for atmospheric extinction, flat-fielded
and calibrated in $T_{\rm mb}$.  Further processing of the data was made with the
CLASS\footnote{CLASS is part of the Grenoble Image and Line Data Analysis
Software (GILDAS), which is provided and actively developed by IRAM, and is
available at http://www.iram.fr/IRAMFR/GILDAS} software.

We also retrieved some high quality  \CII\ data from the SOFIA archive, which
extended the \CII\ map  to the north. This data set was from project 02\_0090
(PI. Els Peeters) and fully calibrated in the SOFIA archive. All data in this paper are presented in main-beam temperature scale.

\begin{figure*}
\centering
\includegraphics[width=7.5cm]{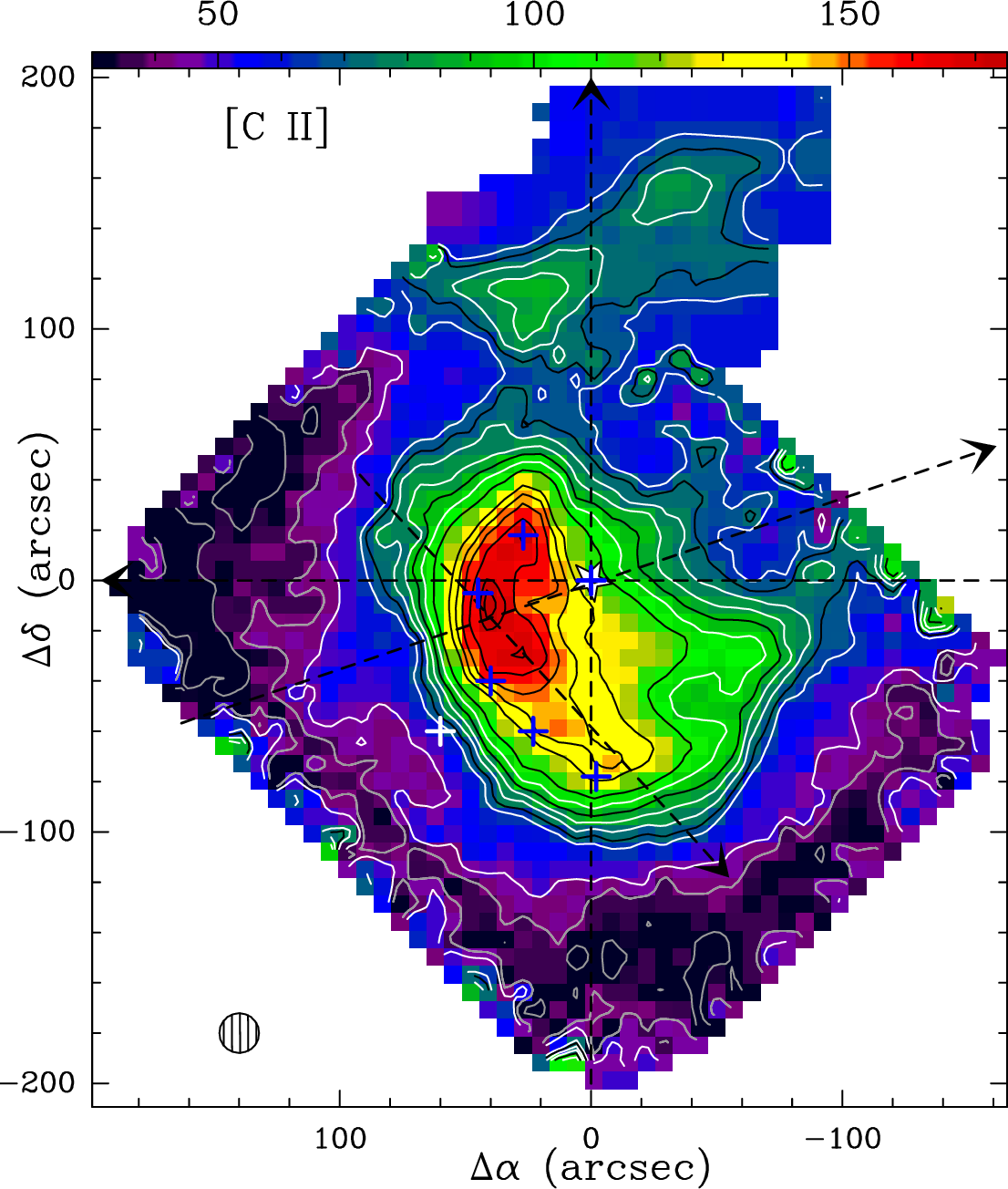}
\includegraphics[width=9.0cm]{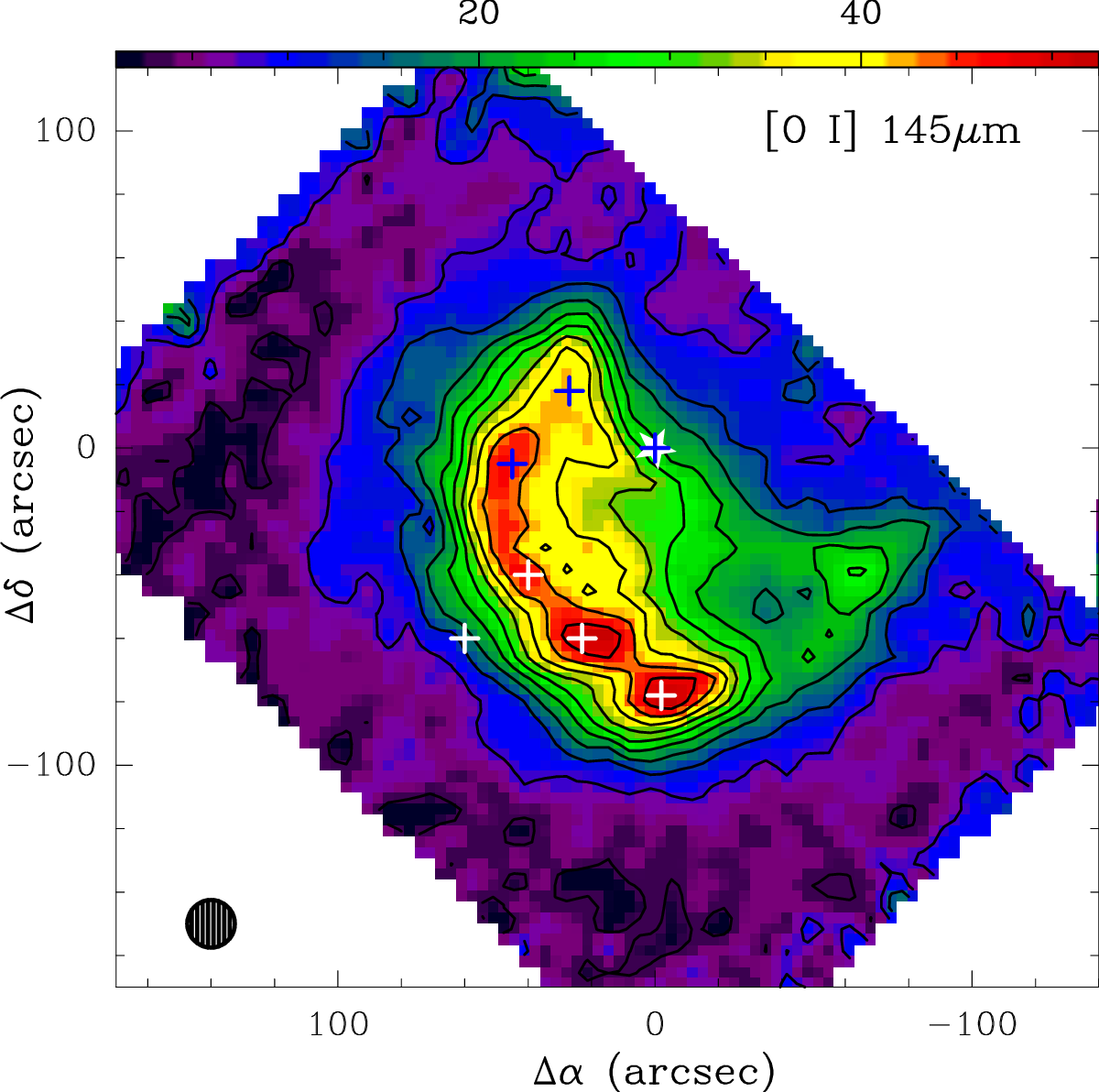} 
\caption{Integrated intensity
images (in both color and contours) of \CII\ at 158\,\micron\ (Left) and \OI\ at
145\,\micron\ (Right). The color scale in K\,\kms\ for each panel shown on the
top. Contour levels (in K\,\kms) for \CII\ are 40, 46, 65, 70, 76, 83, 90, 98,
109, 131, 141, 150, 160, 169 and 173. Contour levels (in K\,\kms) for \OI\ 145
are 3, 5 to 50 in steps of 5. The coordinates are shown as offsets relative to
the center RA: 05$^{\rm h}$41$^{\rm m}$38.39$^{\rm s}$ {\bf Dec:}
-02\arcdeg15\arcmin32\farcs5, which is also the location of the illuminating
star HD\,37903 marked by a star symbol. Positions which are studied in detail
are shown as '+'. The dashed lines drawn in the left panel show the directions along which position-velocity diagrams have been studied in Fig.\ref{fig_allpv}. The beamsizes of the \CII\ and \OI\ 145 maps are 14\farcs1 and 13\arcsec\ respectively and are shown hatched circles in each panel.
\label{fig_intmaps}}
\end{figure*}

\section{Results}

\subsection{Morphology \& Kinematics of the Region}

\begin{figure}
\centering
\includegraphics[width=9.0cm]{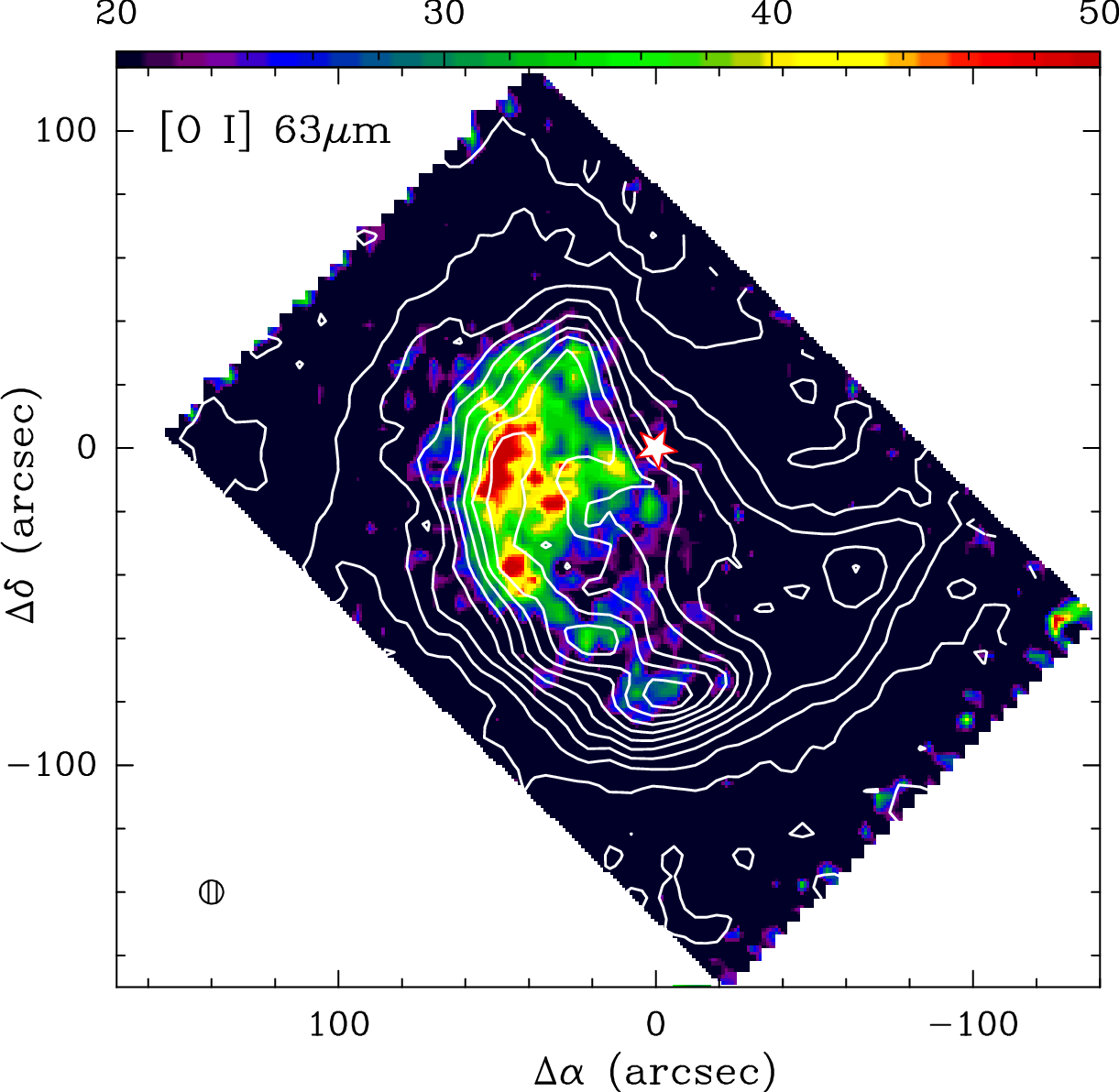}
\caption{Integrated intensity image  (color) of integrated intensities of \OI\
63\,\micron\ overlaid with contours of \OI\ 145\,\micron\ intensities. The color scale is in K\,\kms\ and beamsize for the \OI\ 63\,\micron\ map is shown as hatched circles in the left bottom corner of the figure.  Contour levels (in K\,\kms) for \OI\ 145\,\micron\ 
are 3, 5 to 50 in steps of 5. Rest of the details are the same as in
Fig.\,\ref{fig_intmaps}.
\label{fig_addmap}}
\end{figure}

\begin{figure*}
\centering
\includegraphics[width=6.0cm]{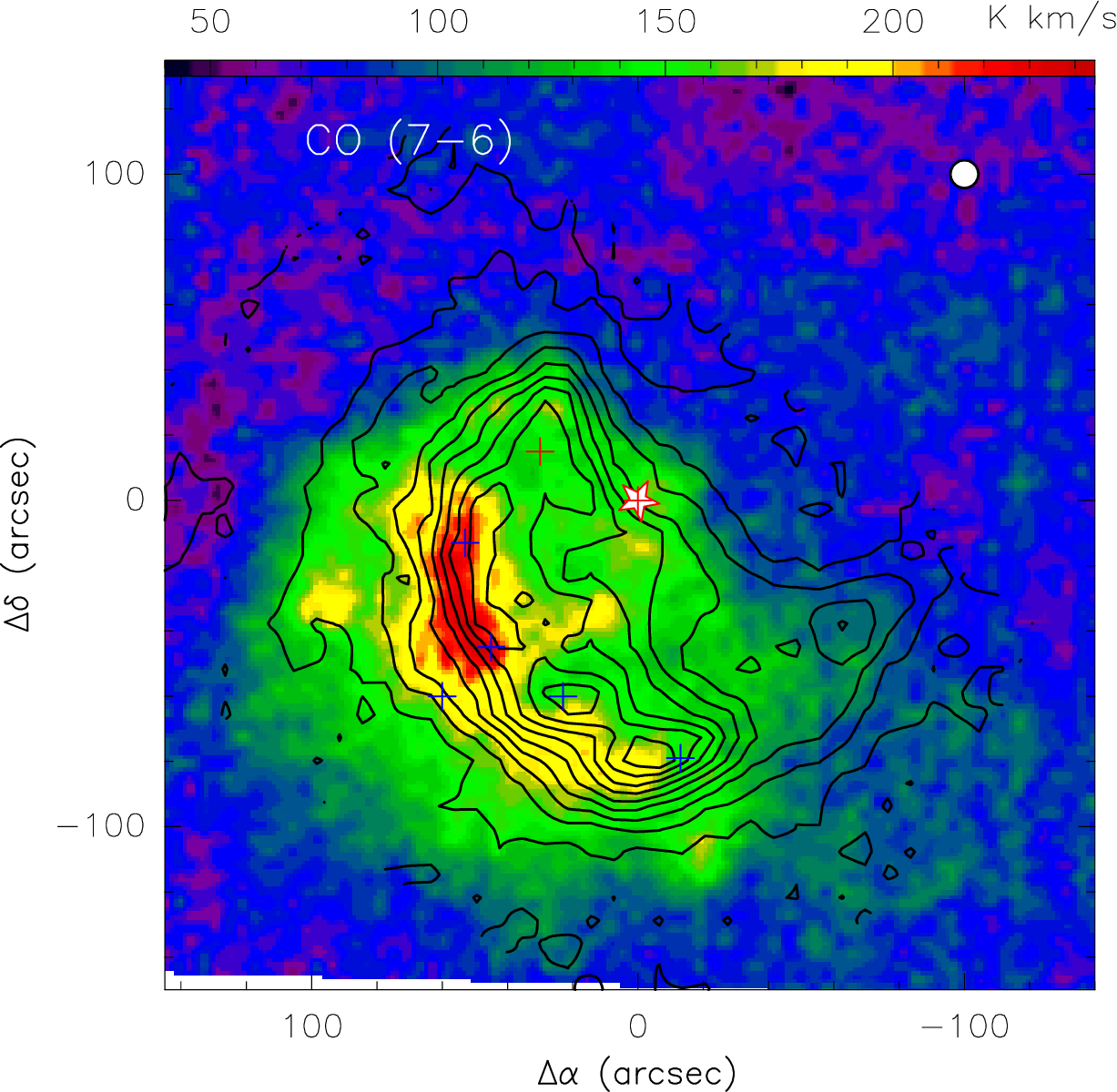}
\includegraphics[width=5.6cm]{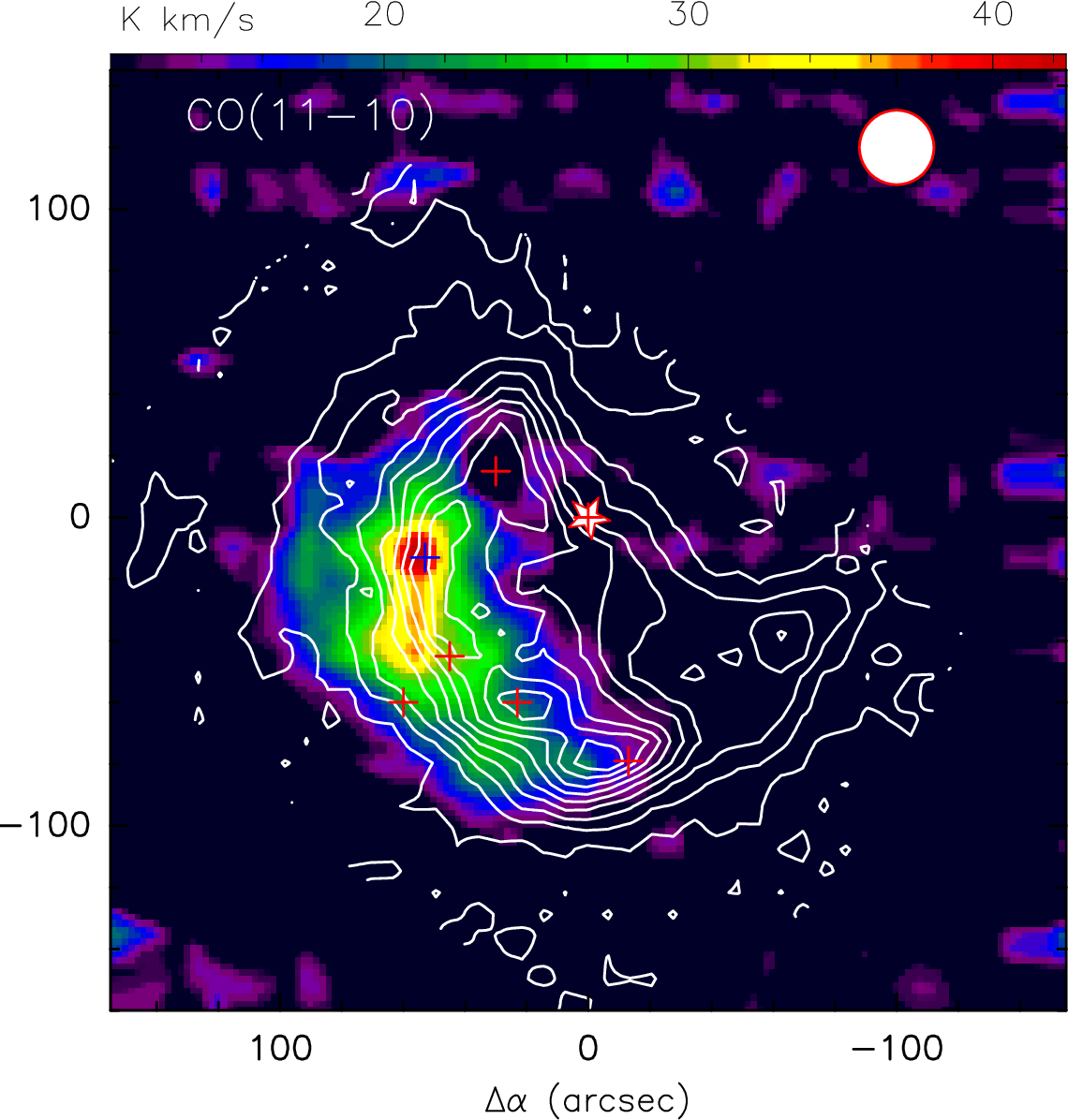}
\includegraphics[width=5.4cm]{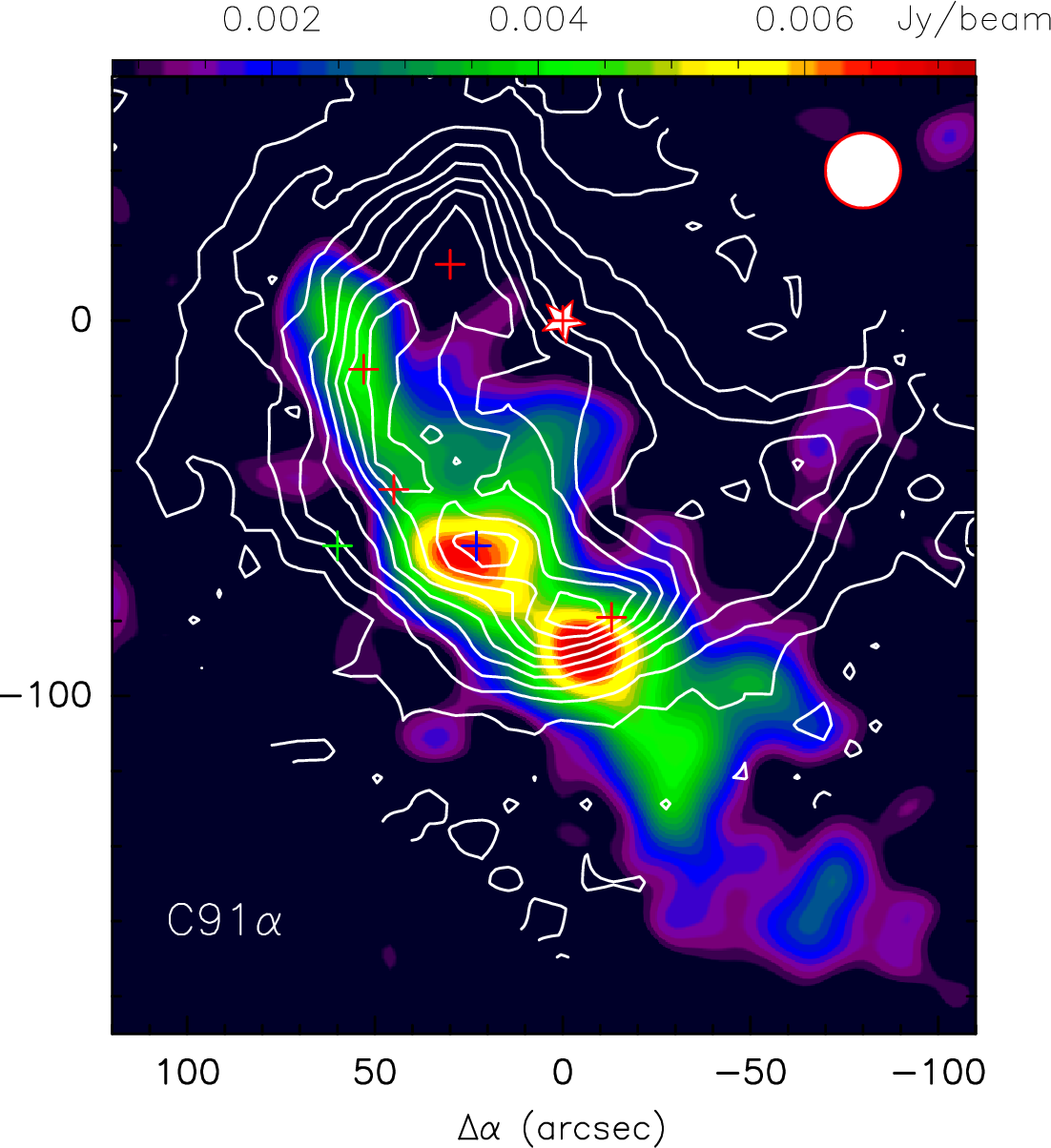}
\caption{Integrated intensity images (color) of CO(7--6), CO(11--10) and
C91$\alpha$ compared with contours of intensities of \OI\ 145\,\micron. The
color scale is in K\,\kms. The beamsizes for the CO(7--6), CO(11--10) and C91$\alpha$ maps  (in color) are 7\farcs7, 23\arcsec\ and 20\arcsec\ respectively and shown on top right corner of 
each panel as a filled white circle. The data for the CO(7--6), CO(11--10) emission observed with APEX and GREAT/SOFIA respectively, are from \citet{Sandell2015} and the C91$\alpha$ map is taken from \citet{Wyrowski2000}.   Contour levels (in K\,\kms) for \OI\ 145 are 3, 5 to 50 in steps of 5. Rest of the details are the same as in Fig.\,\ref{fig_intmaps}. 
\label{fig_over}}
\end{figure*}

\citet{Sandell2015} used maps of multiple CO rotational transitions combined with the \CII\  fine-structure line at 158\,\micron\ to probe the overall morphology of the reflection nebula NGC\,2023. These observations showed that the  \CII\  emission traces an expanding ellipsoidal
shell-like PDR region region, which is surrounded by a hot molecular shell.  In
the southeast, where the \CII\ region expands into a dense, clumpy molecular
cloud ridge, they also detected emission from high-$J$ CO lines, apparently
originating in a thin,  hot molecular shell surrounding the \CII\ emission. These authors found that there was a clear velocity gradient across the nebula, with
the emission being more blue-shifted in the south and southeast and more
red-shifted in the north and northwest, some of which appeared to be due to an
expansion of the nebula. \citeauthor{Sandell2015} also noted that high angular
resolution images of vibrationally excited  H$_2$ and PAH emission suggested
that the  PDR was far from smooth, exhibiting lumpy
ridges and bright filamentary structures. The \CII\ and high-$J$ CO maps looked
smoother, partly because they had insufficient spatial resolution to see such
details. Here we have additionally mapped  the fine-structure transitions of
\OI\  at 63 and 145\,\micron, which are good tracers of the warm and dense
regions of the PDR. The upper energy levels of the 63\,\micron\ and 145\,\micron\ \OI\ transitions are 227.7\,K and 326.6\,K respectively, while formally the critical densities for these transitions are  5\,10$^5$ \cmcub\ and 5\,10$^6$ \cmcub\ respectively, for collisions with H$_2$ \citep{Goldsmith2019}. The  \OI\  63 emission, however, can be very optically thick and self-absorbed, which was seen toward the S1 PDR in RhoOph \citep{Mookerjea2021} and the same is true for NGC\,2023. However, here we have also observed \OI\ 145, which is unaffected by self-absorption. The \OI\ 145 emission is largely optically thin and shows that the dense PDR has a lumpy and filamentary structure.  The \OI\ 145 map
looks qualitatively similar to the \CII\ map (Fig.\,\ref{fig_intmaps}), but the
\OI\ emission is concentrated to the dense south-eastern and southern part of the PDR and more structured, while the \CII\  emission is rather smooth. In the north and northwest the gas densities are too low to collisionally excite \OI. The 
\OI\ 63 map, which has a spatial resolution of 6\ptsec3, shows that the emission is
extremely patchy over the area where it is detected (Fig.\,\ref{fig_addmap}).
The emission is essentially completely self-absorbed in the southeast and south,
while it is seen in emission in the east and northeast, where the absorption
from the foreground cloud is less severe.  Even here it is strongly  affected by
absorption from colder foreground gas giving the integrated line emission a very
patchy and lumpy appearance.


\begin{figure*}
\centering
\includegraphics[width=11.0cm]{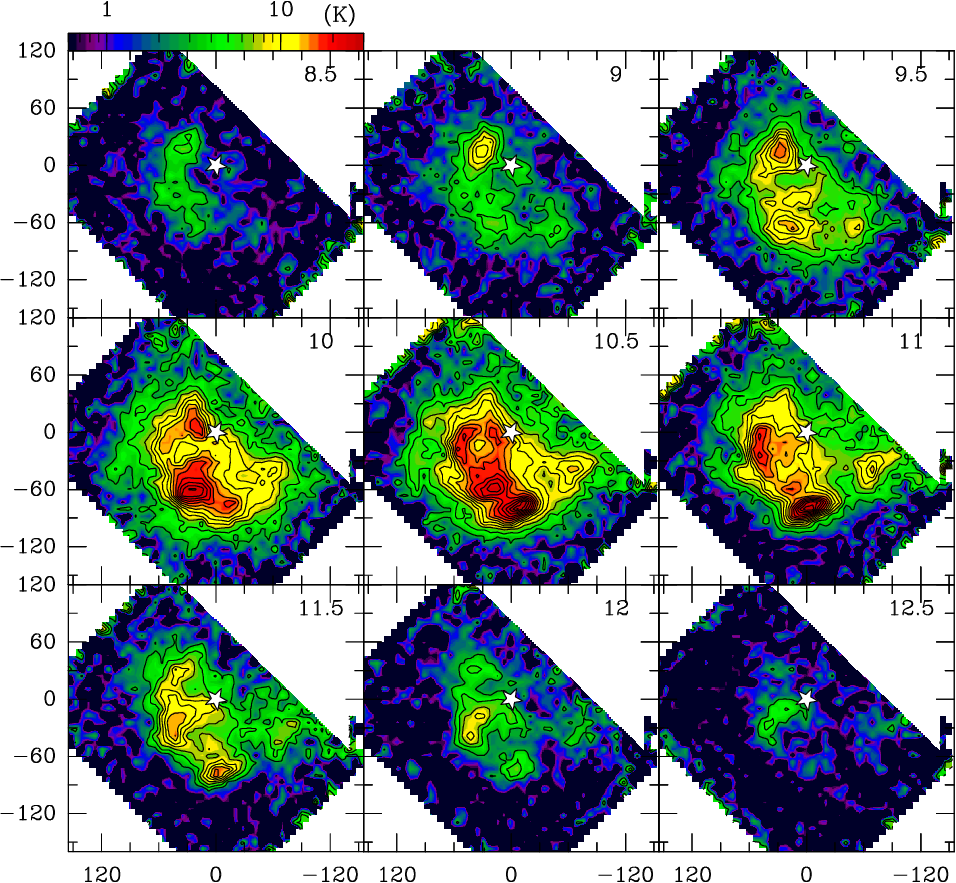}
\includegraphics[width=11.0cm]{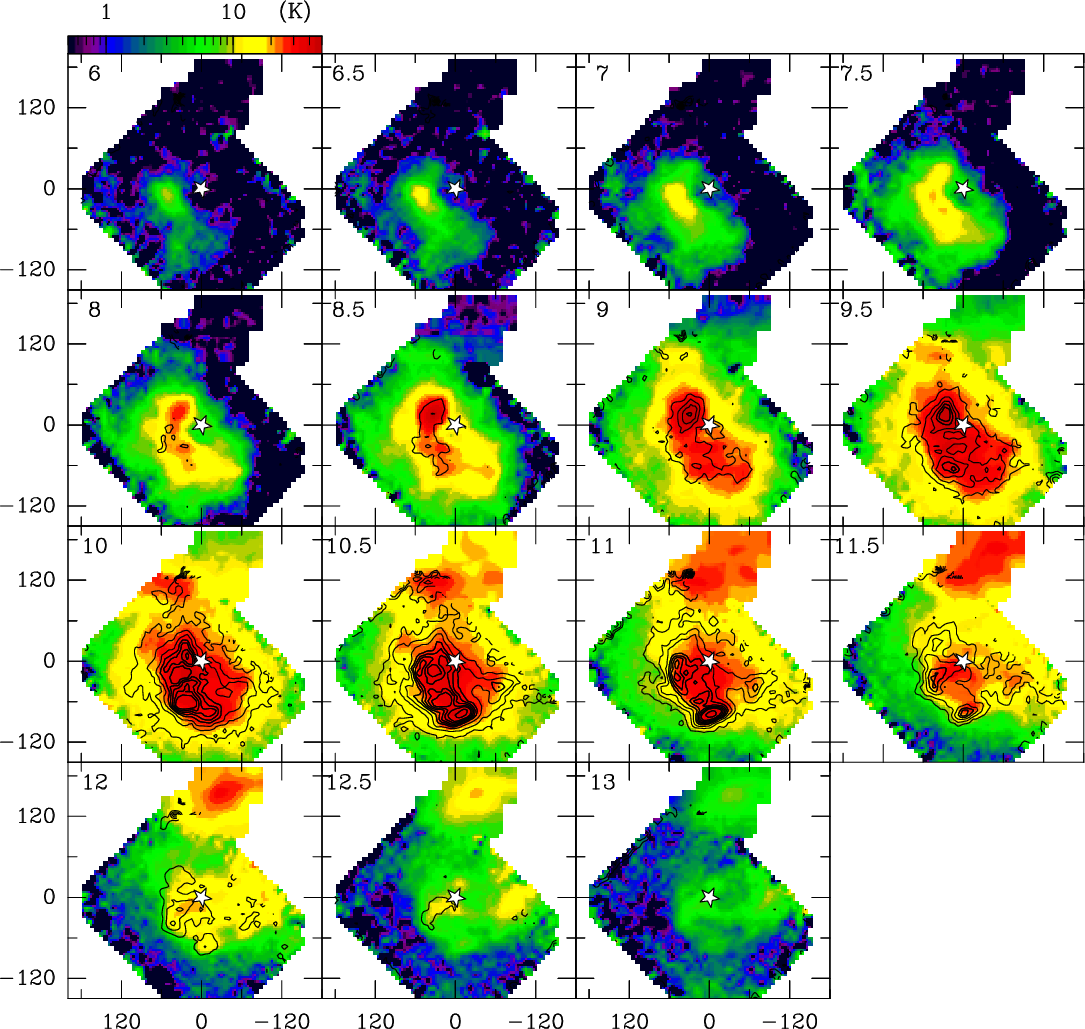}
\caption{{\bf Top panels:} Velocity channel maps for \OI\ 145\,\micron\ in color overlaid with contours at 3 to 30\,K in steps of 2\,K.  {\bf Bottom panels:} Velocity channel maps (in color) for \CII\ overlaid with the \OI\ 145\,\micron\  map in
contours from 3\,K to 30\,K in steps of 2\,K. The color scale for the channel
maps is shown at the top of each set. The position of HD\,37903 is shown as a ``star". The
coordinates in both panels are offsets in RA and Dec relative to the position of
HD\,37903 as in Fig.\,\ref{fig_intmaps}.
\label{fig_oiciichanmap}}
\end{figure*}

 We also compare the \OI\ 145 map with maps of CO(7--6) and CO(11--10)
\citep{Sandell2015}, which also trace the hot, dense PDR as well as with
C91$\alpha$, which was mapped by \citet{Wyrowski2000} (Fig.\,\ref{fig_over}).

The overall morphology of \OI\ 145 matches CO(7--6) and  CO(11--10) quite well. Both appear to trace hot gas in the leading edge of the expanding PDR shell. That CO(7--6) emission is not seen from the entire PDR where \OI\ is detected, is likely due to the fact that the emission is strong enough only from a narrow shell and needs to be viewed almost tangentially to have enough column density to be detected. The emission from the remaining hot gas is likely below the sensitivity of our observations. C91$\alpha$  shows a different morphology than
CO(11--10), although they both appear to originate in  approximately the same
region. However, where C91$\alpha$ is strong, CO(11--10) is faint, like in the
southern part of the PDR shell, where C91$\alpha$ shows two strong peaks. Both
of these peaks are seen in \OI\ 145, which appear to trace the same gas as
C91$\alpha$.  In the east, where C91$\alpha$ is faint there is no clear peak in 
\OI\ 145 either.  We note that although the \OI\ 145 traces the high density ridge
and the clumps in the ridge, the emission is slightly offset compared to
CO(11--10) and C91$\alpha$, which appears to be further outside in the PDR
shell, i.e., closer to the cold molecular cloud (Fig.\,\ref{fig_over}).  Based
on the detailed analysis of the high-$J$ CO lines, \citet{Sandell2015} estimated
the hot molecular shell to be between 90--120\,K and have  densities $n\sim
10^5$--10$^6$\,\cmcub.

The velocity-channel map of \OI\ 145 emission clearly show that the dense PDR
clumps along the ridge are also at different velocities with the south-western
clumps being more red-shifted than the north-eastern clump
(Fig.\,\ref{fig_oiciichanmap}).  At v = 9\,\kms\ the region northeast of  HD\,37903
dominates the \OI\ 145 and  \CII\ emission. This emission peak, at $\sim$
+27\arcsec, +18\arcsec, coincides with the strongest \CII\ emission peak, $\sim$
55 K, in the whole map. It is also seen as a distinct peak in CO(6--5), CO(7--6)
and CO(11--10), but it is relatively faint compared to the emission at $\sim$ v =
10.5\,\kms, which dominates the emission in the southeastern part of the PDR.
This suggests that the northeastern emission peak may have somewhat lower gas
densities than the PDR emission in  the ridge. 

\begin{figure*}
\centering
\includegraphics[width=3.95cm]{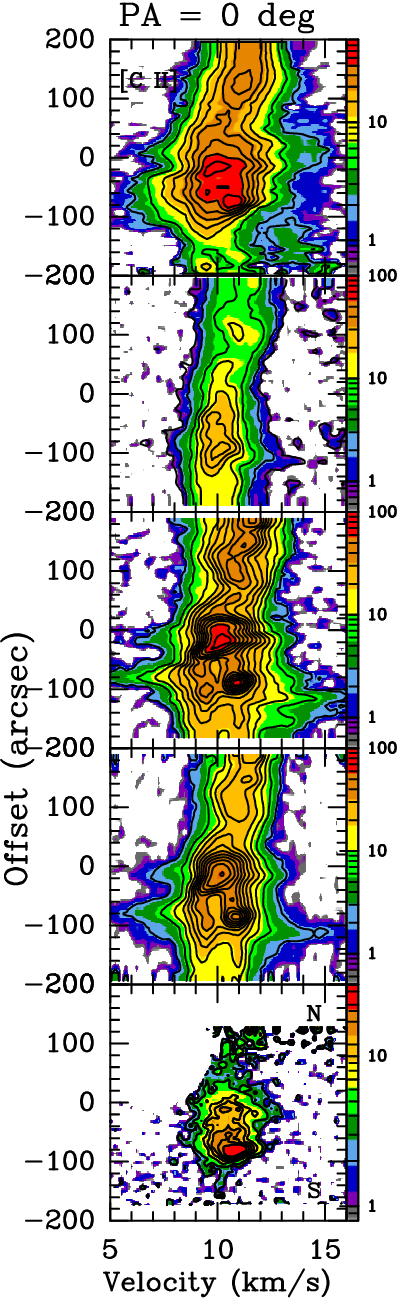}\hspace{0.4cm}
\includegraphics[width=3.65cm]{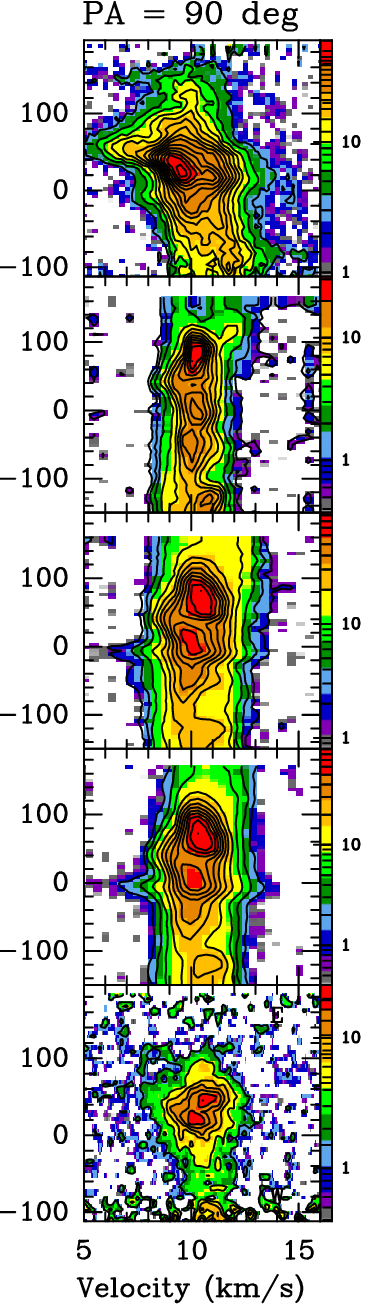}\hspace{0.4cm}
\includegraphics[width=3.7cm]{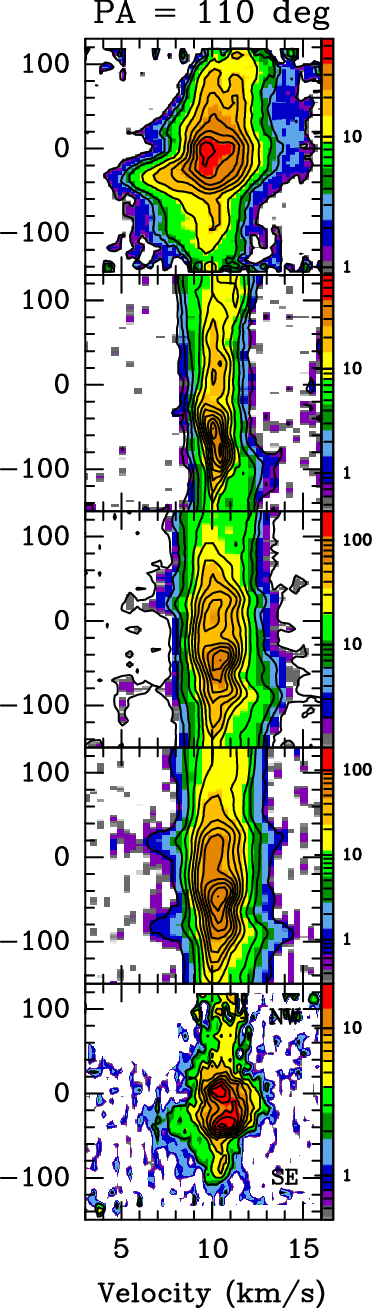}\hspace{0.4cm}
\includegraphics[width=3.58cm]{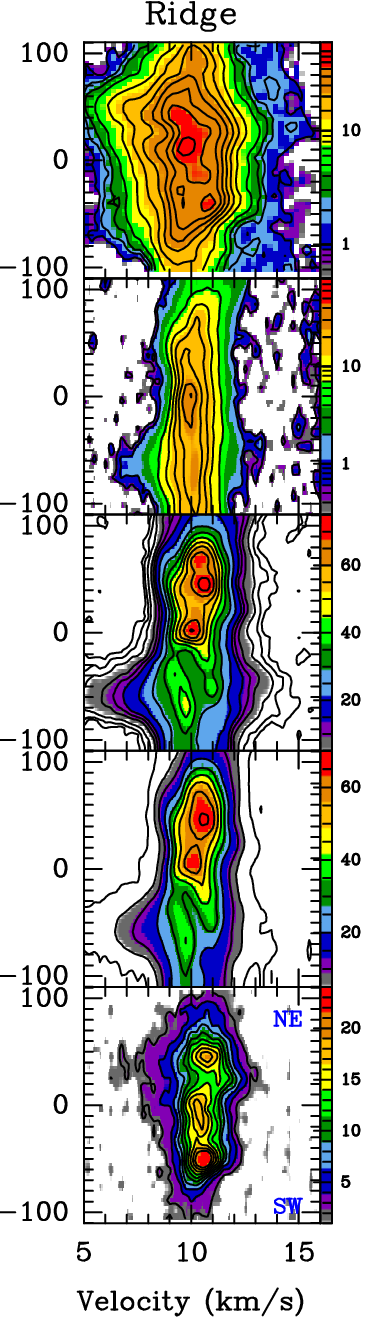}
\caption{Position-velocity diagrams for \CII, $^{13}$CO(3--2), CO(4--3),  CO(6--5) and
\OI\ 145 (from top to bottom in each column) along directions given by position
angles PA of 0, 90\arcdeg\ and 110\arcdeg\ as well as along the ridge. The position angles are measured counterclockwise from north. The colorscales (in K) are shown next to each panel. All the plots are at their respective original resolutions.
\label{fig_allpv}}
\end{figure*}

Optical images show that the NGC\,2023 reflection nebula is quite large with a
size of $\sim$ 10\arcmin\ $\times$ 10\arcmin. In the near- and mid-IR it is only
about half the size, $\sim$ 5\farcm 2 $\times$ 5\farcm3
\citep{Mookerjea2009}. This is the regime where the emission is dominated by the
PDR. In the near-IR one can see that HD\,37903 illuminates a bright ridge $\sim$
120\arcsec\ -- 150\arcsec\ to the north, which intercepts some of the FUV
radiation. This ridge is dominated by fluorescent H$_2$ emission, see e.g,  the
narrowband  H$_2$ imaging by \citet{Field1998}. The feature
\citeauthor{Field1998} calls the IR triangle is quite bright in \CII\
(Fig.~\ref{fig_intmaps}). The peak of the \CII\ emission (+27\arcsec,
+18\arcsec) coincides with the head of what \citeauthor{Field1998} refer to as the
Seahorse. It is clear that the PDR shell opens up to the north, as well as in  a
region  east of the IR triangle, Fig.~\ref{fig_intmaps}. Even in the southeast,
where the \cplus\ region is surrounded by the dense molecular cloud, one can
see that FUV radiation escapes, like the tongue of \CII\ emission sticking out
at  a position angle (measured counter-clockwise from North) of $\sim$ 110\degr\ (Fig.~\ref{fig_intmaps}).

The position-velocity diagrams, Fig.\,\ref{fig_allpv}, provides us with an alternative way
to examine the morphology of the PDR and what the different PDR tracers show us.
There is a clear velocity gradient from south to north with the emission in the
northern part of the PDR being more red-shifted, seen in all the tracers, even
\OI\ 145, although the  \OI\ 145 map only partially covers the PDR ridge. This
suggest that the IR triangle and the northern ridge is on the backside of the
PDR. Near HD\,37903 the \CII\ emission shows both blue-and red-shifted emission,
suggesting that the whole \cplus\ region is embedded in the cloud, because
one sees photo evaporation flows both from the front and the back side of the
PDR. The CO(4--3) and CO(6--5) show an outflow just south of HD\,37903 and strong
blue-and red-shifted emission, where the position-velocity diagram crosses the outflows from Sellgren C and D. Although \CII\ also shows strong blue-shifted emission in the south, it
is almost certainly due to strong photo evaporation, as one can see from the
velocity-channel maps (Fig.~\ref{fig_oiciichanmap}). These maps show strong
blue-shifted emission over the whole PDR in the southeast. The east-west
position-velocity diagram shows the strong blue-shifted photo-evaporation flow
inside the PDR shell and it appears that some FUV radiation ``leaks'' through
the shell, i.e., one can see \CII\ emission at distances $>$ 100\arcsec\ from
HD\,37903 at roughly the same velocity, $\sim$ 9.9 \kms. Closer to the star the
\CII\ emission appears to be somewhat self-absorbed. \OI\ 145 shows blue-shifted
emission east of HD\,37903, indicating that the photo evaporating gas is so
dense that it is seen even in \OI\ 145.  To the west, the \OI\ 145 becomes more
red-shifted, which is not really noticeable in \CII. The CO lines pick up the
blue-shifted outflow slightly west of HD\,37903. The position-velocity diagram
at PA 110\degr\ shows one of the regions where \CII\ breaks through the PDR
shell.  However,  it looks about the same as in the east-west cut. Towards the
west-northwest  the \CII\ splits up into two velocity components. However, both
 \OI\ 145 and $^{13}$CO(3--2) only show a single velocity component roughly halfway 
 between the two \CII\ peaks, indicating that \CII\ must be self-absorbed.

The position-velocity diagram along the ridge cuts through the two strong 
C91$\alpha$ peaks. The southern one is faint in \CII\, while it is quite prominent 
in \OI\ 145. Here the emission is shifted toward $\sim$ 10.7\,\kms, whereas the
clump to the northeast of it is $\sim$ 10.2,\kms{} (Fig.\,\ref{fig_allpv}). The velocity
of the \OI\ 145 emission appears to follow that of C91$\alpha$  \citep{Wyrowski2000},
while these emission clumps are unnoticeable  in \CII. The CO lines show strong
blue-shifted emission from a molecular outflow powered by the young star Sellgren
D.

\subsection{Spectral Profiles of \OI\ lines}

\begin{figure*}
\centering
\includegraphics[width=8.6cm]{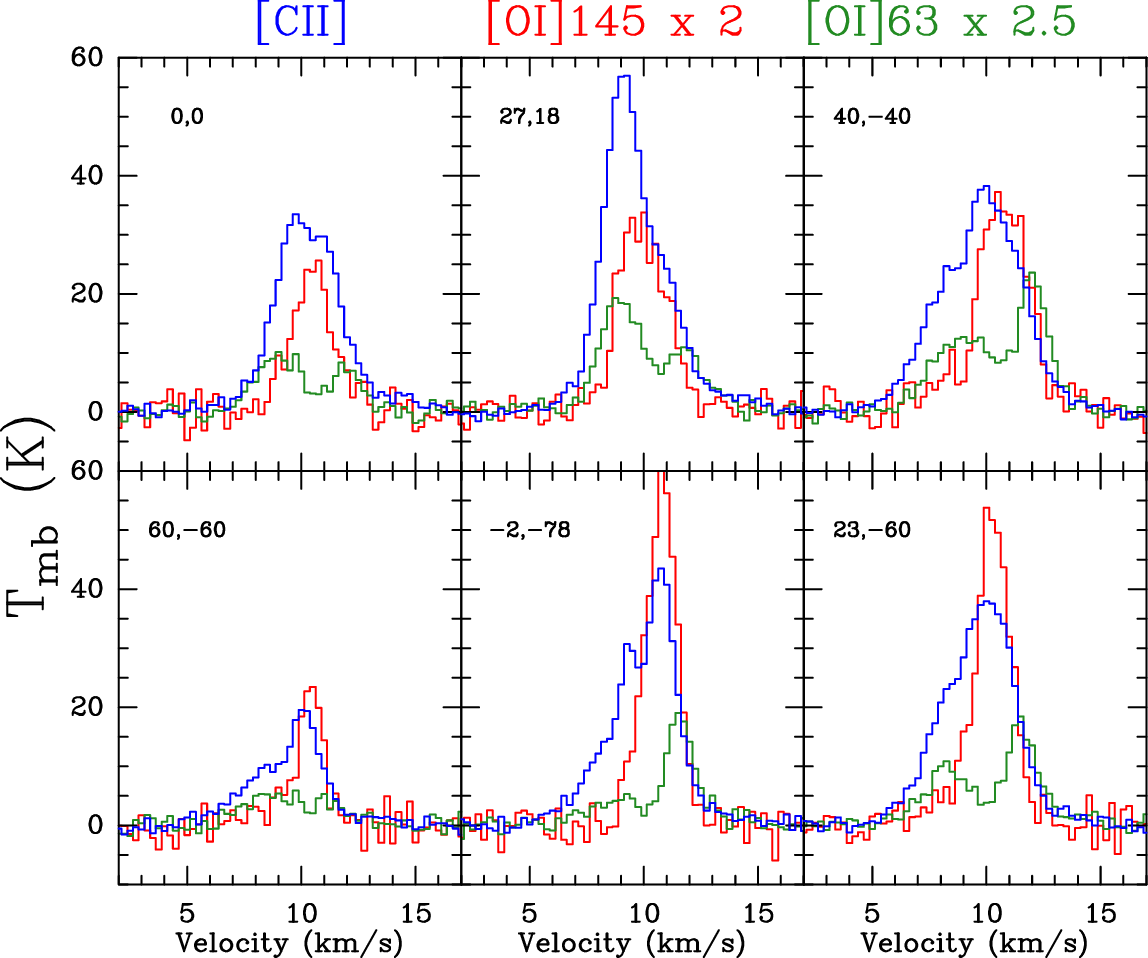}
\includegraphics[width=8.1cm]{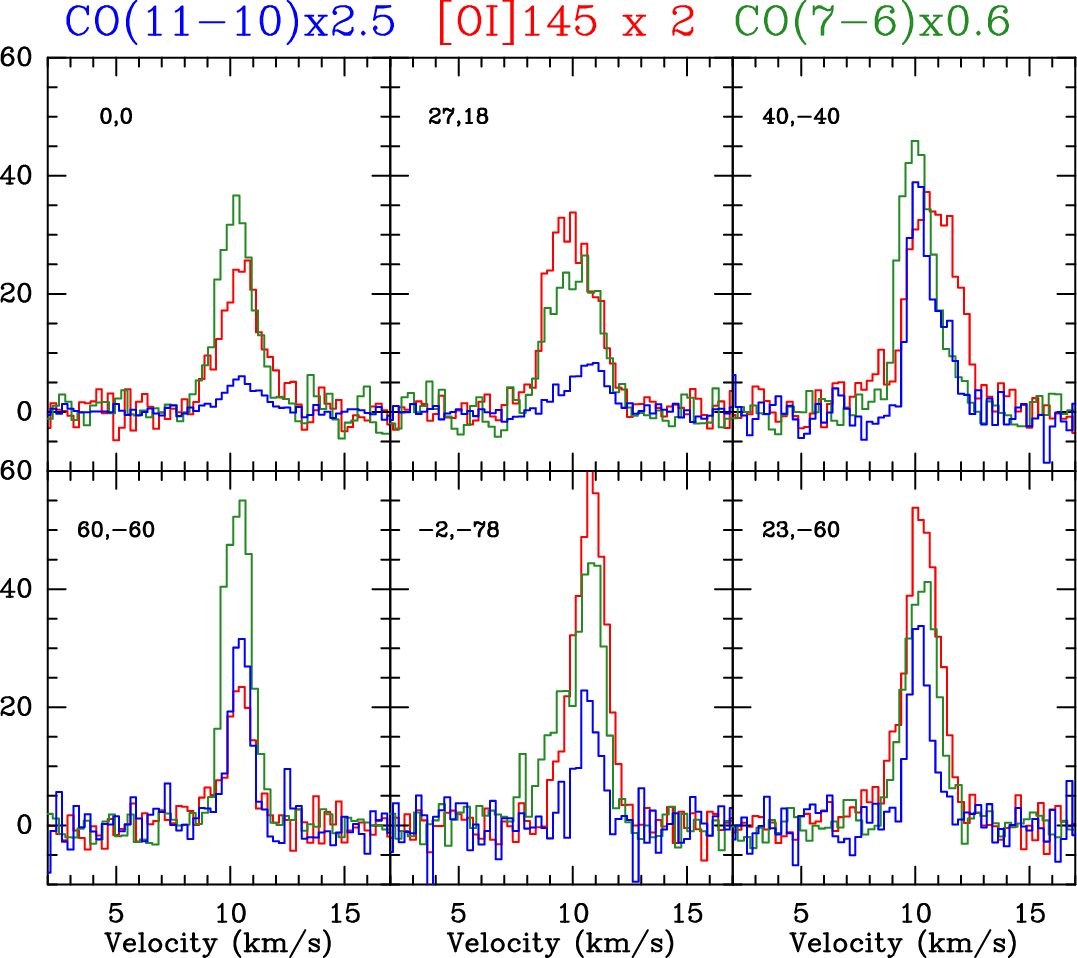}
\caption{Comparison of the \OI\ 145\,\micron\ spectra at selected positions in the
NGC\,2023 region (marked in Fig.\,\ref{fig_intmaps}) with spectra of \CII, \OI\
at 63\,\micron, CO(11--10) and CO(7--6). As indicated the spectra are scaled by
arbitrary factors for better visibility. The temperature scale is in $T_{\rm
mb}$(K). The CO(7--6) and CO(11--10) are from \citet{Sandell2015}. The CO(11--10) spectra are at the native resolution of 23\arcsec, all other spectra are at a common resolution of 15\,\arcsec.
\label{fig_spec}}
\end{figure*}

Based on the model proposed by \citep{Sandell2015} the spectral profiles of CO
and \CII\ detected NGC\,2023 consist of several velocity components:
the quiescent cloud seen in low J CO lines and  $^{13}$CO, the PDR, and 
line wings from high velocity molecular outflows seen in the outskirt of the reflection 
nebula.  The \CII\ spectra also show prominent blue or red-shifted wings from photo evaporation flows.  In order to see which of these components contribute to the \OI\ emissions, we have
selected six positions, some of which already  were analyzed by \citep{Sandell2015}.
All  spectra are smoothed to a common angular resolution of  15\arcsec\
(Fig.\,\ref{fig_spec}). We have also fitted the \OI\ 145 spectra using Gaussian
profiles consisting of one or two velocity components as appropriate and
compared the fitted parameters with those derived for \CII\ (this work) and
CO(6--5), (7--6) and (11--10) transitions (Table\,\ref{tab_gaussfit}). We have
not used  \OI\ 63\,\micron, because the line is strongly self-absorbed over the
whole area that we have mapped. The \OI\ 145 spectra are all 
single-peaked, although we do see clear blue-shifted line wings in several positions, like  at 40\arcsec, -40\arcsec\  and  23\arcsec, -60\arcsec, where \CII\ shows very strong blueshifted wings. The position 23\arcsec, -60\arcsec, which coincides with the C91$\alpha$ emission peak 2,  also shows  blue-shifted emission in \OI\ 145, and even in C91$\alpha$. At  position 60\arcsec, -60\arcsec, the \CII\ line is strongly self-absorbed at the velocity
of \OI\ 145, making the blue-shifted line wing appear stronger than the emission from the PDR (Table~\ref{tab_gaussfit}). \citet{Sandell2015} interpreted the double-peaked \CII\ spectra that they saw in their \CII\ map as emission coming from the front and the backside of the PDR shell, but the optically thin \OI\ 145 spectra show that the dip in the \CII\ spectra is due to 
self-absorption, not two separate velocity components. Neither can  we distinguish more
than one velocity component in the CO spectra, although there are areas, see Fig. ~\ref{fig_spec} and Table~\ref{tab_gaussfit}, where the lines appear broadened, most likely due to contribution both from the front and the back wall of the PDR,

 In most of  the positions the \OI\ 63 spectra show wings and line-widths  which are comparable to the \CII\ spectra, although they are often completely self-absorbed at the center of the line.  The \OI\ 145 spectra are much narrower than the \OI\ 63\,\micron\ and show profiles which
are almost identical to the CO(7--6) emission, except at position -2\arcsec, -78\arcsec\
where the CO(7--6) shows strong blue-shifted from the outflow powered by Sellgren's star C (Sandell, priv. communication).



\begin{table*}
\caption{Gaussian fits to spectra at selected positions.
\label{tab_gaussfit}}
\begin{tabular}{llrrrrrr}
\hline\hline
Offset & Tracer & $\int T_{\rm mb} dv$\phantom{0} & $V_{\rm LSR}$\phantom{00} & $\Delta V$\phantom{00}
&  $\int T_{\rm mb} dv$\phantom{0} & $V_{\rm LSR}$\phantom{00} & $\Delta V$\phantom{00}\\
(\arcsec,\arcsec) && (K\,\kms) & (\kms{}) & (\kms) & (K\,\kms) & (\kms) & (\kms)
\\
\hline
(0.0,0.0)   & \CII{}$^a$       & 113.7$\pm$1.2 & 10.24$\pm$0.02 & 3.22$\pm$0.04&&&\\
 & $^{13}$CO(3--2)   &  32.0$\pm$0.4& 10.24$\pm$0.01 &  1.61$\pm$0.02 &&&\\ 
           & \OI\ 145  & 27.2$\pm$1.2  & 10.48$\pm$0.04 &  2.16$\pm$0.10 & & & \\
         &   CO(6--5)        &  132.0$\pm$0.5 &  10.25$\pm$0.01 &  2.38$\pm$0.01 &&& \\
         &   CO(7--6)        &   85.0$\pm$0.8 &  10.24$\pm$0.01 &  2.02$\pm$0.02 &&& \\
         &   CO(11--10)    & 4.4$\pm$0.2 & 10.51$\pm$0.05 & 1.91$\pm$0.13 &&&\\

\hline

(27,18)      &   \CII{}$^b$           & 159.7$\pm$3.9& 9.10$\pm$0.02 &  2.02$\pm$0.04 & 51.5$\pm$4.1 & 11.19$\pm$0.06 & 1.97$\pm$0.12\\
          & $^{13}$CO(3--2)   &  32.9$\pm$0.5 & 10.11$\pm$0.02 &  2.35$\pm$0.04 &  &  & \\
           & \OI\ 145                & 43.2$\pm$1.3 & 9.89$\pm$0.04 & 2.50$\pm$0.09 & & & \\
         &   CO(6--5)        &  128.6$\pm$0.3 &  10.05$\pm$0.01 &  2.77$\pm$0.01 &  &   & \\ 
         &   CO(7--6)        &  80.3$\pm$1.0 &  10.07$\pm$0.01&  2.51$\pm$0.03 &  & &  \\  
         &   CO(11--10)      & 7.0$\pm$0.4 &  10.63$\pm$0.06 & 2.10$\pm$0.15 &  &  & \\

\hline
(40,-40)          &   \CII\           & 72.8$\pm$0.0 & 10.73$\pm$0.01 &  2.80$\pm$0.03 & 79.0$\pm$0.5 & 8.73$\pm$0.0.01 & 3.58$\pm$0.02\\
 & $^{13}$CO(3--2)   &  53.5$\pm$0.4 & 10.06$\pm$0.01 &  1.44$\pm$0.01 & & & \\
         & \OI\ 145           & 48.2$\pm$1.6& 10.60$\pm$0.04 &2.46$\pm$0.09  & 5.6$\pm$1.4 &  7.71$\pm$0.43& 3.00$\pm$0.00$^c$\\
         &   CO(6--5)        &  165.2$\pm$0.3 & 10.30$\pm$0.01 &  2.13$\pm$0.01&  &   &  \\
         &   CO(7--6)        &  116.2$\pm$0.7&  10.24$\pm$0.01 &  1.95$\pm$0.01 &  &  & \\
         &   CO(11--10)      & 28.7$\pm$0.3&  10.29$\pm$0.81 & 1.61$\pm$0.02 &  &  & \\       
           \hline

(60,-60)      &   \CII\           & 39.6$\pm$1.1 & 10.08$\pm$0.04 &  2.20$\pm$0.00$^c$ &  15.2$\pm$1.4 & 7.65$\pm$0.09 & 2.40$\pm$0.24\\
          & $^{13}$CO(3--2)   &  47.7$\pm$0.4 & 10.22$\pm$0.01 &  1.09$\pm$0.01 & & &\\
         & \OI\ 145    & 10.6$\pm$0.9 & 10.41$\pm$0.05 & 1.13$\pm$0.12 &
          5.8$\pm$2.8 & 8.87$\pm$1.04 & 3.00$\pm$0.00$^c$\\
        &   CO(6--5)        &133.9$\pm$0.3 &  10.37$\pm$0.01 &  1.58$\pm$0.01\\
         &   CO(7--6)        &102.8$\pm$0.7 &  10.35$\pm$0.01 &  1.47$\pm$0.01\\
         &   CO(11--10)     & 18.1$\pm$0.9 &  10.49$\pm$0.03 & 1.13$\pm$0.07\\
\hline
(-2,-78)  &   \CII\   & 126.5$\pm$1.1 & 10.17$\pm$0.01&  2.87$\pm$0.03 & &  &  \\
        &    $^{13}$CO(3--2)   &  55.5$\pm$0.5 & 10.11$\pm$0.01 &  1.98$\pm$0.02 & &&\\  
         & \OI\ 145        &50.2$\pm$1.1  & 10.77$\pm$0.02 & 1.64$\pm$0.04\\   
         &   CO(6--5)$^a$        &  87.0$\pm$0.3 &  10.81$\pm$0.01 &  1.56$\pm$0.01 & & &\\
         &   CO(7--6)$^b$       & 68.6$\pm$0.8 &  10.68$\pm$0.01 & 1.51$\pm$0.02& &  &\\
         &   CO(11--10)    & 14.2$\pm$1.2 & 10.70$\pm$0.07 & 1.49$\pm$0.15\\

\hline
(23,-60)         &   \CII\           & 112.4$\pm$2.7 & 10.27$\pm$0.03 & 2.49$\pm$0.09  & 50.9$\pm$1.8 &  8.19$\pm$0.05 &  2.42$\pm$0.17\\
         & $^{13}$CO(3--2)   &  49.3$\pm$0.9 &  10.00$\pm$0.01 &  1.32$\pm$0.01 &&&\\
         & \OI\ 145           & 40.7$\pm$3.8  & 10.33$\pm$0.02 & 1.61$\pm$0.08 & 14.7$\pm$3.9 & 9.50$\pm$0.38  & 4.30$\pm$0.65 \\
         & CO(6--5)        &  120.1$\pm$0.0 &  10.23$\pm$0.01 &  1.97$\pm$0.01 & &&\\
         &   CO(7--6)        &  90.6$\pm$0.8 &  10.21$\pm$0.01 &  1.75$\pm$0.02 &&&\\
         &   CO(11--10)      & 12.6$\pm$0.8 & 10.19$\pm$0.04 & 1.12$\pm$0.10 &&&\\
\hline
\end{tabular}

\hspace*{-5.8cm}{\bf Note:} The CO(7--6) and CO(6--5)  spectra have been convolved to a resolution of 15\arcsec. \\ \hspace*{-2.8cm}The \thCO(3--2) and CO(11--10) spectra are at their respective original resolutions of 18\farcs5 and 23\arcsec.\\
\hspace*{-8cm}$^a$ Both blue- and red-shifted outflow wings masked out \\ 
\hspace*{-9.8cm}$^b$ Blue-shifted outflow wing masked out \\
\hspace*{-11cm}$^c$ Not fitted, i.e., kept constant
\end{table*}

\subsection{Column Density of the Atomic Oxygen}
\label{sect_twoslab}
\begin{figure}
\includegraphics[width=8.5cm]{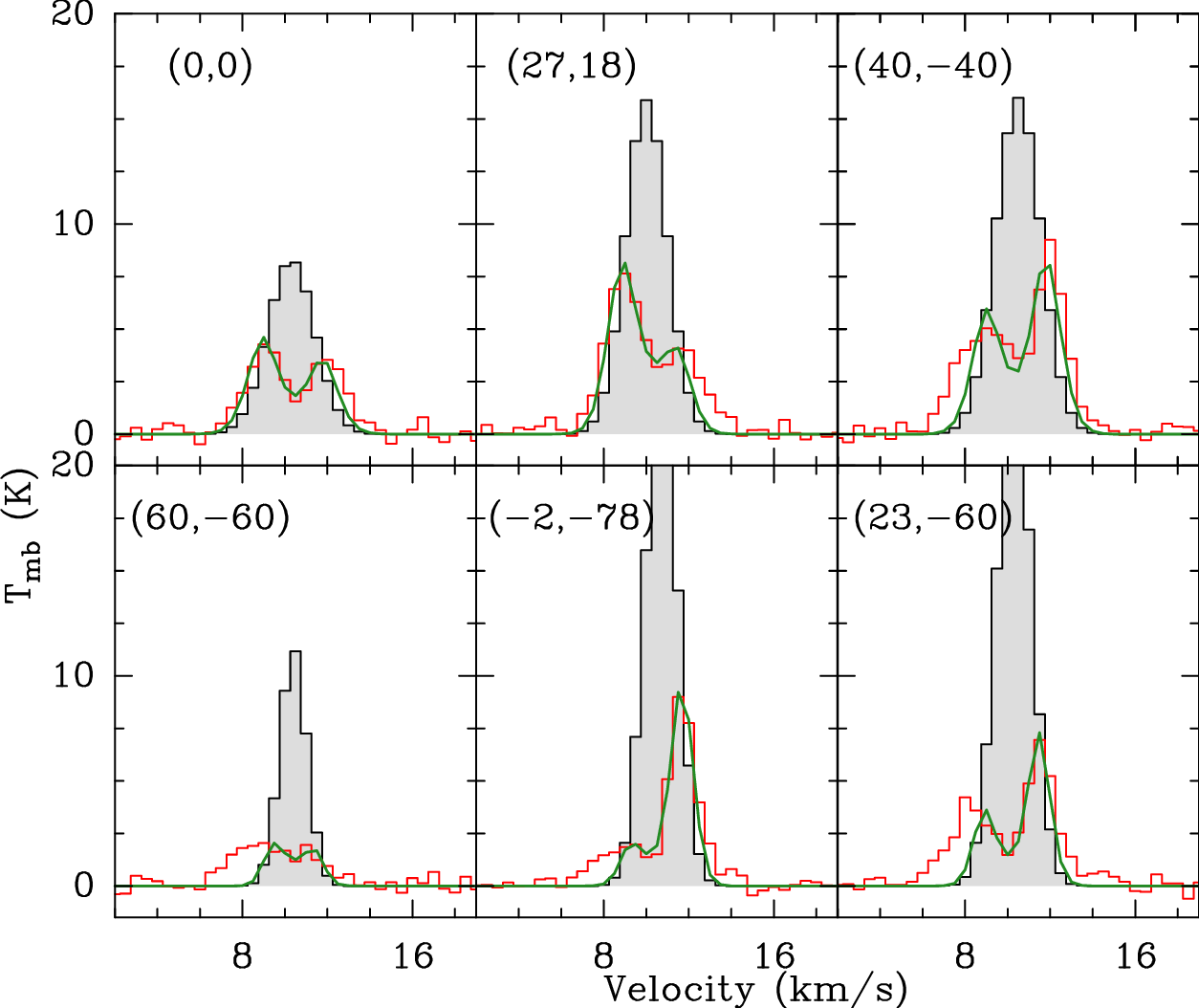}
\caption{The smooth curve (green) shows the fit to the \OI63\ spectrum (red)
obtained by attenuating the scaled (by a factor 2 corresponding to the typical
intensity ratio between the two \OI\ lines in temperature units) \OI\ 145\,\micron\
spectrum (filled grey histogram) by absorption due to foreground material. 
\label{fig_oi63fit}}
\end{figure}
\begin{figure}
\centering
\includegraphics[width=8.5cm]{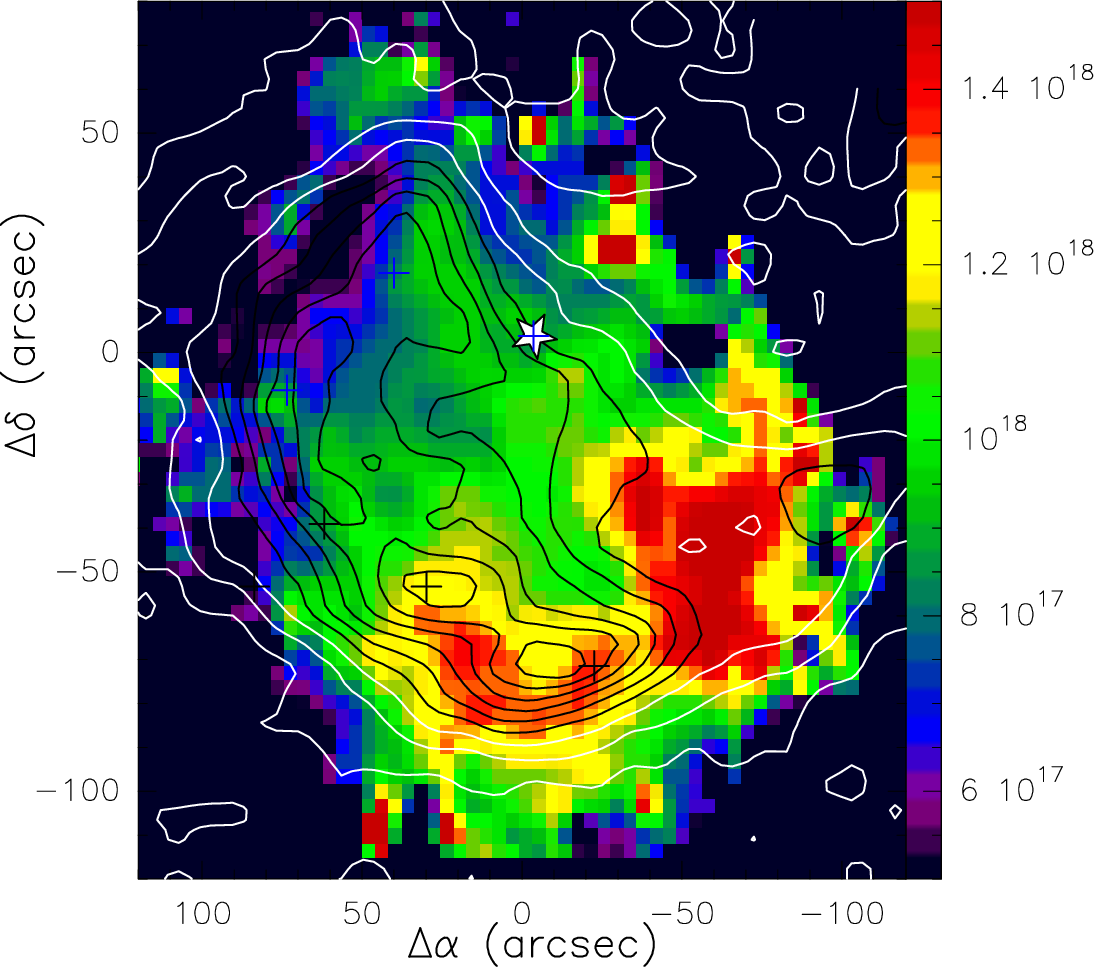}
\caption{The column density distribution of foreground PDR gas responsible for
the \OI\ 63 absorption features in the NGC\,2023 region plotted in color and
overlaid with the dense PDR gas traced by \OI\ 145 emission. The contour levels
in K\,\kms for the \OI\ 145 are 3, 5 to 50 in steps of 5. This image shows that
the densest, most opaque area of the foreground cloud is in the south and
southwest, while the foreground extinction is much lower in the east and in the
north.
\label{fig_o63abs}}
\end{figure}

\begin{table*}
\begin{center}
    
\caption{Peak optical depth, corresponding velocity, and the FWHM
of the foreground absorbing gas derived from modeling the 
\OI\ 63\,\micron\ spectra with a two-slab model. Column density of oxygen in foreground cloud ($N_{\rm abs}$) is estimated using equation (1) and in the background PDR ($N_{\rm em}$) estimated using RADEX (Fig.\,\ref{fig_radex}). 
\label{tab_twoslab}}
\begin{tabular}{crrrccccc}
\hline
Position & $\tau_0$ & $v_{\rm LSR}$ & $\Delta v$ & $N_{\rm abs}$(O) & $N_{\rm em}$(O)$^a$ & $T_{\rm kin}^b$(\cplus) & $N$(\cplus) & $N$(H$_2$)$^c$\\
& & \kms & \kms & (10$^{18}$ \cmsq & (10$^{18}$) \cmsq) & K & (10$^{18}$ \cmsq) & (10$^{21}$ \cmsq) \\
\hline
\hline
(0, 0)    & 2.2  & 10.4  &  2.0 & 0.88 & 3.4(2.8) & 74 & 3.4 & 7.8\\ 
(27,18)   & 2.2  & 10.3 & 2.2 & 0.97 & 6.6(5.5) & 107 & 3.6 & 9.1\\ 
(40,-40)  & 2.4 &  10.4  &  2.0 & 0.96 & 7.3(6.0) &  80 & 3.8 & 22.9\\ 
( 60,-60) & 2.9  & 10.4  & 1.8  & 1.0 & 2.0(1.6) & 60 & 2.4 & 23.9\\ 
( -2,-78) & 3.3 &  10.3  &  1.9 & 1.3 & 7.0(5.7) & 84 & 3.2 & 23.0\\ 
( 23,-60) &  3.5 &  10.1  &  1.9 & 1.3 & 9.1(7.2) & 81 & 3.8 & 29.0\\ 
\hline
\hline
\end{tabular}
\end{center}

\hspace{-1.0cm}$^a$ Estimated from integrated \OI\ 145\ intensity (Table\,\ref{tab_gaussfit}) assuming $T_{\rm kin}$ = 120\,K and $n$(H$_2$)= 2\,10$^5$ \& 10$^6$\,\cmcub. \\
\hspace{-5.8cm}The number in brackets corresponds to $N$(O) for $n$(H$_2$)=10$^6$\,\cmcub.\\
\hspace{-4.0cm}$^b$ Estimated from Planck-corrected peak of the \CII\ spectrum as described in Sec. 4\\
\hspace{-0.5cm}$^c$ Estimated from pixel-by-pixel grey-body fitting of 160, 250, 350 and 500\,\micron\  dust continuum emission using {\em hires}, an improved\\
algorithm for the derivation of high-resolution surface densities from multiwavelength far-infrared Herschel images \citep{Menshchikov2021}. \\
\end{table*}

The \OI\ 63\,\micron\ emission is strongly self-absorbed over most of the
reflection nebula, whereas the \OI\ 145 is always observed in emission with no evidence
for self-absorption. Even though \CII\  and \OI\ 63\,\micron\ shows bright wings
due to photo-evaporation flows, the photo-evaporating gas is not dense enough to
be seen in  \OI\ 145  except in a few areas, see e.g., spectra at positions
40\arcsec, -40\arcsec, and 23\arcsec, -60\arcsec, where  \OI\ 145 shows a clear
blue-shifted line wing. In order to derive the distribution of atomic oxygen
based on the 63 and 145\,\micron\ spectra, we fit the \OI\ 63 spectra using a
two-slab toy model: the hot PDR  shell (lying on either side of HD37903 along
the line-of-sight) is assumed to be the background layer, while  the colder
oxygen on the front side is taken to be the foreground layer. While the gas
in this layer, where the PDR merges into the cold molecular cloud is largely
molecular, most of the oxygen is still atomic. We further assume that the
emission from the background layer is captured by the \OI\ 145 emission.  In this
case the colder foreground oxygen primarily  attenuates the background spectral
emission at 63\,\micron\ by the factor $\exp\left[-\tau_0 \exp\left[-4 \ln 2 (v
- v_0)^2/\Delta v^2\right]\right]$. Here $\tau_0$, $v_0$, and $\Delta v$ denote
the peak optical depth, the velocity at which $\tau_0$ of the foreground cloud
occurs and the full width at half maximum (FWHM) of the absorption profile due
to the foreground material. As pointed out earlier, the \OI\ 63 emission arising
from a transition involving lower energy levels also trace the blue- and
red-shifted gas in  photo-evaporation flows. The \OI\ 145 emission, which is only
excited in the warm and dense gas, does not show these broad line wings. The
background \OI\ 63 emission is modeled by Gaussian profiles generated from fits
to the observed \OI\ 145 spectrum scaled by a factor of 2,  which corresponds to the
typical intensity ratio between the two \OI\ lines in temperature units and is equal to $I_{63\mu m}/I_{145\mu m}$ = 25 in energy units. Calculations for warm, medium density PDRs with \tkin = 100\,K and $n$=10$^5$\,\cmcub\ suggest a $I_{63\mu m}/I_{145\mu m}$ ratio 
between 20 to 33 \citep[Fig.10][]{Goldsmith2019}. Assumption of a higher ratio of 33 
leads to an increase in the derived oxygen column density by 10--20\% of the value obtained in this work. Figure\,\ref{fig_oi63fit} shows an example of
the results of the fits to the \OI\ 63 using the two-slab toy model at the
selected positions and Table\,\ref{tab_twoslab} presents the corresponding
fitted parameters. At the selected positions, the central velocity of the
absorbing foreground cloud lies between 10.1--10.7\,\kms\ which matches well
with the velocity of the \OI\ 145 emission spectra and have FWHM between
1.8--2.4\,\kms.  This suggests that the \OI\ 63 absorption occurs in the colder
foreground layers of the same PDR gas that accounts for the emission of \OI\ 145.
We have performed the fitting of the \OI\ 63 profile over a larger number of
positions in the map and obtained the distribution of the foreground cold
absorbing component (Fig.\,\ref{fig_o63abs}).

The fits using the \OI\ 145 spectra as the model for the background component do not
reproduce the red- and blue-shifted wings observed in the \OI\ 63 spectra, since the gas 
in the photo-evaporation flows causing these wings is not dense enough to excite \OI\ 145. 
We estimate the column density of
the absorbing foreground layer by assuming that all the oxygen atoms reside in
the lowest level of their ground state. The column density of all O atoms in the
foreground layer can be  then estimated from the center opacity $\tau_0$ and
width (in \kms) of the \OI\ 63 line by

\begin{equation}
    N({\rm O^{0}}) = 2\times 10^{17} \tau_0 \Delta v ~~~~~ {\rm cm^{-2}}
\end{equation}

The estimated column densities for the foreground gas ranges from 8\,10$^{17}$ to 2\,$\times 10^{18}$\,\cmsq. We can estimate a lower limit to the extinction if we assume that all the oxygen is atomic. With  a typical gas abundance of 3.2$\times 10^{-4}$ for oxygen and assuming most of the hydrogen to be molecular we estimate $N$(H$_2$) of the foreground gas to range between  1.2--3.2\,10$^{21}$\,\cmsq, which translates to a minimum A$_{\rm V}$ of 1.3.  However, since one third of the oxygen is tied up in CO in molecular gas a more realistic estimate for the A$_{\rm V}$ is 2 mag.

 For the estimated column densities, based on non-LTE calculations using RADEX
 \citep{VdTak2007}, the \OI\ 145 line is optically thin over a large range of
 temperatures (20--300\,K) and volume densities (10$^4$--10$^7$\,\cmcub). This is
 also consistent with our observations. Since the $T_{\rm mb}$ ratio for the two
 \OI\ lines depends on the physical conditions, the derived values of the peak
 optical depth are subject to the ratio assumed in this work. However, the
 fitted central velocity and line widths for the absorption profile are fairly
 robust against the assumed scale factor. Furthermore, in the estimate of $N$(O)
 we have assumed that all O atoms in the absorbing layer are in the ground
 $^3$P$_2$ level, which is reasonable because  there is no hint of absorption 
 in \OI\ 145. 
 

Since the \OI\ 63\,\micron\ is  optically thick it cannot  be used to estimate
the column density of oxygen atoms in the background layer. We thus use the
integrated intensities of the \OI\ 145\,\micron\ spectral lines to estimate
$N$(O). For this we need an estimate of the temperature and density of the
emitting gas.  Since the \OI\ 145 emission seems to follow the CO(11--10)
emission quite closely, we can use the temperature and density estimated for 
CO(11--10), for which  \citet{Sandell2015} estimated a  kinetic temperature of
120\,K and densities of 2\,$10^5$--1\,$10^{6}$\,\cmcub. Table\,\ref{tab_twoslab}
presents the column density of atomic oxygen at the seven selected positions
that is required to produce the observed \OI\ 145 emission
(Table\,\ref{tab_gaussfit}), estimated based on non-LTE calculations using RADEX
\citep{VdTak2007} for these kinetic temperatures and densities
(Table\,\ref{tab_twoslab}). Figure\,\ref{fig_radex} shows a plot of the
intensity of \OI\ 145 line as a function of $N$(O) for $T_{\rm k}$ =
120\,K  and $n$(H$_2$) = (0.2--1)\,10$^6$\,\cmcub.

\subsection{The morphology of the reflection nebula and the PDR illuminated by HD\,37903}

\citet{Sandell2015} proposed that the PDR illuminated by HD\,37903 formed an
expanding thin, hot shell which is roughly spheroidal with the south-eastern
part being almost spherical because the expansion is slowed down by the dense
surrounding molecular cloud and the shell being much more extended to the
north-west. They interpreted the double peaked \CII\ spectra seen on the
north-western side of the star as originating from the front and backside of the
PDR shell. They argued that the reason for not seeing clear double split spectra on the southeastern side, was because the expansion was slowed down by the dense
molecular cloud. While their model appears reasonable, we now know that it has
some serious flaws. Even though we do see double peaked \CII\ spectra north-west
of HD\,37903, see e.g. Fig.~\ref{fig_allpv}, \OI\ 145 is single peaked and
centered between the two  \CII\ peaks. This confirms  that what we see is
self-absorption in \CII, not two velocity components. The \CII\ lines are quite
broad, so it is probably true that they are broadened by emission from both the
front and the back side of the PDR, but they cannot be separated into two
distinct velocity components. The foreground extinction is definitely lower in the
north and northwest. We do see some blue-shifted \CII\ wings due to photo-evaporation flows. This  is consistent with the cloud emission being more redshifted in both  $^{13}$CO(3--2) and \CII\ indicating that it is behind the PDR. It appears more likely that the PDR is fan-shaped rather than 'egg-shaped' and that it is most likely open in the north and north-west allowing FUV photons
to escape, hence creating a large reflection nebula dominating the emission north and north-west of HD\,37903.

\section{Discussion}

Analysis of the observed \OI\ 63 and \OI\ 145 spectra suggests that the  strong absorption features seen in the \OI\ 63 profiles are caused by colder atomic oxygen  lying between the observer and the warm PDR emission excited by  HD\,37903. The \CII\ spectra are also affected by self-absorption, but it is far less severe and barely noticeable over most of the reflection nebula. At the peak (27\arcsec, 18\arcsec) of the \CII\ emission  (Fig.\,\ref{fig_spec}), the self-absorption results in the shifting of the \CII\ spectra toward  blue-shifted velocities compared to \OI\ 145 and the high-$J$ CO lines, which are unaffected by self-absorption. Even the \CII\ spectrum on  HD\,37903 appears to be somewhat affected by self-absorption.

The two-slab toy model, which we used in Section~\ref{sect_twoslab} to model the observed \OI\ 63 absorption features, shows that the column density of atomic oxygen in the foreground absorbing gas is quite substantial, $\sim$ 10$^{18}$~\cmsq. This is because most of the oxygen remains atomic way into the cloud, i.e. from the hot PDR layer up to about an A$_V$ of 10, as already shown by \citet{Tielens1985}, while most of the carbon becomes molecular, i.e., gets tied up in CO. 
The contribution to the oxygen column density from the surface of the molecular cloud, which only gets ionized by the general interstellar radiation that is a factor of 10$^4$ or
10$^5$ times weaker than the PDR illuminated by HD37903, is negligible.
The estimated \OI\ column densities depend to some extent on  the ``model" for the background emission, which have been constructed using the \OI\ 145 spectra, but as we can see in Table~\ref{tab_twoslab} the computed column densities only depend weakly on the assumed gas densities.

The \OI\ 63 spectral profiles seen in NGC\,2023 are similar to those observed in other strongly self-absorbed regions like S1 in RhoOph \citep{Mookerjea2021} and toward W3 \citep{Goldsmith2021},  where  the  absorption was estimated to be due to colder foreground  gas with $N$(O) of 3\,10$^{18}$ and 2--7\,$10^{18}$\,\cmsq\ respectively.  The method for estimating the background  for the S1 PDR in Rho\,Oph  was similar to what we used in this paper, while for  W3, which was not observed in \OI\ 145  it was instead estimated by fitting the \OI\ 63 spectra for positions which showed almost no evidence for absorption. In the \OI\ 63 map in NGC\,2023 there are only a few positions which show a single emission peak without any trace of absorption, thus the approach to use the \OI\ 145 emission which has been measured is more robust.

Since we detect the \thCII\ $F$ = 2--1 line in essentially the entire area mapped in \CII, it is possible to check whether the \CII\ emission is optically thick although the \CII\ map does not go deep enough to measure  \CII\ to \thCII\ ratio in individual positions. We have therefore divided up the map into smaller regions  of about 1 sq. arcmin, each containing about 50--70 spectra. When we did this we found that the ratio of intensities between \CII\ to \thCII\ within errors is $\sim$ 30--40 after correcting for the relative intensity of the $F$ = 2--1 line, 0.625. 
Considering the isotope ratio $^{12}$C/$^{13}$C of 70 for the Orion region, this suggests the \CII\ optical depth to be $\sim$ 1.5--2, consistent with the value of 2 derived by \citet{Sandell2015}. We obtain an estimate of the minimum kinetic temperature $T_{\rm kin}$(\cplus) from the Planck-corrected peak of the \CII\ spectrum  assuming a beam filling factor of unity. The lower limit of $T_{\rm kin}$(\cplus) thus estimated for the entire map ranges between 60--92\,K. Assuming an optical depth of 2 and $T_{\rm kin}$(\cplus), and the integrated \CII\ intensity to estimate $N$(\cplus) using the Equation (26) from \citet{Goldsmith2012} modified for optically thick emission as follows:


{\small
\begin{equation}
N(C^+) = {\rm 2.91\times 10^{15}\left[1+0.5{\rm
e^{91.25/T_{kin}}}\left(1+\frac{A_{ul}}{C_{ul}}\right)\right] \frac{\tau}{1-e^{-\tau}} \int T_{\rm mb}d\upsilon}
\end{equation}
}
where $A_{\rm ul}$ = 2.3$\times 10^{-6}$\,s$^{-1}$, \tkin\ is the gas kinetic temperature, the collision rate is $C_{\rm ul}$ = $R_{\rm ul}n$ with $R_{\rm ul}$ being the collision rate coefficient with H$_2$ or H$^0$, which depends on $T_{\rm kin}$, and $n$ is the volume density of H. Since the critical density of the \CII\ transition is  $n_{\rm cr}$=3000\,\cmcub, and since it is likely that most of  the  \CII\ detected could be at such densities along with some emission arising from clumps with densities exceeding 10$^5$\,\cmcub, we assume a density of  10$^4$\,\cmcub\ to estimate the $N$(\cplus) density distribution of the region. For this calculation we use the kinetic temperatures of \cplus\ estimated as above, a density of $n$=10$^4$\,\cmcub, and excitation from  \cplus--H$_2$ collisions, with  $R_{ul}$ = 3.8$\times 10^{-10}$\,cm$^3$\,s$^{-1}$. The value of $N$(\cplus) estimated in the region show a small range of values between (1.8--4.4)$\times 10^{18}$\,\cmsq. We note that barring a few exceptions \citep{Okada2015,Mookerjea2021} for the estimate of $N$(\cplus) often a kinetic temperature of 100--120\,K is assumed in contrast to using an estimate from the observed \CII\ peak temperatures as we have done here. From equation (2) we find that for a temperature of 60\,K the $N$(\cplus) estimated will be approximately twice the value that would be estimated for $T_{\rm kin}$=120\,K. As seen at the selected positions, throughout the entire region the ratio of column densities of \cplus\ and O atoms (seen in emission) is  consistent with the comparable to the solar [O]/[C] abundance ratio of 3.5 within the uncertainties (Table\,\ref{tab_twoslab}). This is also consistent with the structure of the PDR expected where the \OI\ 145 emission arises from a dense hot region which also emits in \CII, although \CII\ emission additionally arises from the more diffuse and colder PDR gas.

Based on the far-infrared dust continuum maps at 160 to 500\,\micron\ we estimate the molecular hydrogen column density to be around 7$\times 10^{21}$\,\cmsq, while to the south, beyond the ridge and into the molecular cloud the values are typically $\sim 2 \times 10^{22}$\,\cmsq. The estimated  H$_2$ and \cplus\ column densities at the selected positions are consistent with the solar C/H abundance ratio of 3$\times 10^{-4}$ (Table\,\ref{tab_twoslab}).


\section{Summary \& Conclusions}

We have studied the geometry of the NGC\,2023 reflection nebula using spectrally resolved observations of the 63 and 145\,\micron\ transitions of \OI\ along with the \CII\ 158\,\micron\ transition which enable us to map the three-dimensional distribution of the photodissociated gas vis-a-vis the location of the illuminating source HD\,37903. Combination of the velocity-resolved \OI\ 145 spectra with the \CII\ spectra have allowed us to improve our understanding of  the morphology of the region.  We conclude that the PDR is fan-shaped rather than 'egg-shaped' and that it is most likely open in the north and north-west allowing FUV photons to escape, that creates a large reflection nebula dominating the emission north and north-west of HD\,37903. The \CII\ emission detects PDR gas uniformly distributed with $N$(\cplus)$\sim 0.9$--$2.2\,10^{18}$\,\cmsq\ suggest the contribution of both diffuse and dense components, while the \OI\ 145 emission detects dense PDR gas with $N$(O) between (2--10)\,10$^{18}$\,\cmsq. Additionally, the self-absorbed profile of the \OI\ 63 indicate the presence of cold low-excitation atomic oxygen with $N$(O)$\sim $0.8--1.5\,$10^{18}$\,\cmsq\ associated with the nebula and lying in the foreground between the illuminating source and the observer that is most pronounced towards the southern and south-western part of the nebula. The self-absorbed \OI\ 63 profiles observed in NGC\,2023 together with similar results in other Galactic PDRs \citep{Mookerjea2021,Goldsmith2021} suggest the need to exercise caution  when modelling galactic and extragalactic \OI\ 63 intensities, in particular if spectrally unresolved, without accompanying observations of the optically thin \OI\ 145 line.

\section*{Acknowledgements}
BM acknowledges the support of the Department of Atomic Energy, Government of India, under Project Identification No. RTI 4002. This research has made use of the VizieR catalogue access tool, CDS, Strasbourg, France (DOI : 10.26093/cds/vizier). The original description of the VizieR service was published in 2000, A\&AS 143, 23"



\section*{Data Availability}

The new data observed with SOFIA that is presented here will be available from the SOFIA data archive (https://irsa.ipac.caltech.edu/applications/sofia/).

\bibliographystyle{mnras}
\bibliography{n2023} 

\appendix
\section{Result of non-LTE calculations using RADEX}

\begin{figure}
\centering
\includegraphics[width=8cm]{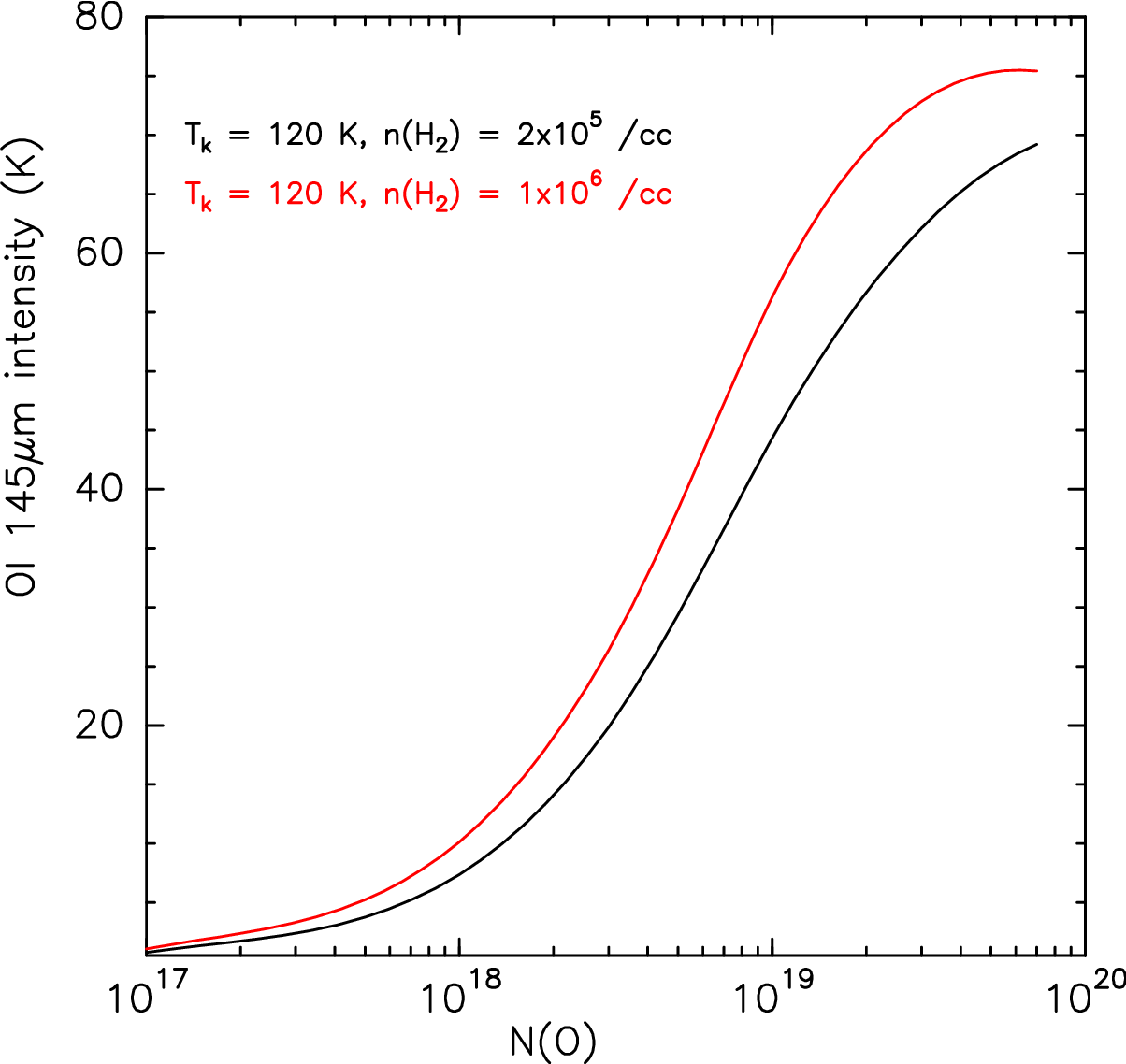}
\caption{The peak \OI\ 145 intensities for $\Delta v = 2$\,\kms, $T_{\rm kin}$=120\,K and $n$(H$_2$) = 2\,10$^5$ (black) and 10$^6$\,\cmcub (red) calculated using the non-LTE radiative transfer models RADEX \citep{VdTak2007} and using collision rate coefficients from \citep{Lique2018}. 
\label{fig_radex}}
\end{figure}

Figure\,\ref{fig_radex} shows the result of the non-LTE radiative transfer model calculations to estimate $N$(O) assuming \tkin = 120\,K and $n$(H$_2$) =  2\,10$^5$ and 10$^6$\,\cmcub\ performed using RADEX \citep{VdTak2007}. 

\label{lastpage}
\end{document}